\documentclass[preprint,preprintnumbers,amsmath,amssymb,superscriptaddress]{revtex4}

\usepackage{bm}
\usepackage{graphicx}
\usepackage{subfig}
\usepackage{array}
\usepackage{color}
\usepackage[section]{placeins}
\allowdisplaybreaks[1]

\makeatletter

\newcommand{\Rmnum}[1]{\expandafter\@slowromancap\romannumeral #1@}
\makeatother

\begin{document}

\title{$B_{(s)} \rightarrow D^{**}_{(s)}$ form factors in HQEFT and model independent analysis of relevant semileptonic decays with NP effects\footnote{This work was supported  in part by the National Natural Science
Foundation of China under Grant No. 12147214.}}

\author{Ya-Bing Zuo}\email{zuoyabing@lnnu.edu.cn}\quad
\author{Hong-Yao Jin} \quad
\author{Jing-Ying Tian} \quad
\author{Jia Yi} \quad
\author{Han-Yu Gong} \quad
\author{Ting-Ting Pan}
\affiliation{Department of physics, Liaoning
Normal University, Dalian 116029, P.R.China }
\affiliation{Center for Theoretical and Experimental High Energy Physics, Liaoning Normal University, Dalian 116029, P.R.China}

\begin{abstract}

The form factors of $B_{(s)}$ decays into P-wave excited charmed mesons (including $D^*_0(2300)$, $D_1(2430)$, $D_1(2420)$, $D^*_2(2460)$ and their strange counterparts, denoted generically as $D^{**}_{(s)}$) are systematically calculated via
QCD sum rules in the framework of heavy quark
effective field theory (HQEFT). We consider contributions up to the next leading order of heavy quark expansion and
give all the relevant form factors, including the scalar and tensor ones only relevant for possible new physics effects. The expressions for the
form factors in terms of several universal wave functions are derived via heavy quark expansion. These universal functions can be evaluated through QCD sum rules. Then, the numerical results of the form factors are presented. With the form factors given here, a model independent analysis of relevant semileptonic decays $B_{(s)} \rightarrow D^{**}_{(s)} l \bar{\nu}_l$ is performed, including the contributions from possible new physics effects. Our predictions for the differential decay widths, branching fractions and ratios of branching fractions $R(D^{**}_{(s)})$ may be tested in more precise experiments in the future.

\noindent
{\bf Keywords:} excited charmed meson, form factor, HQEFT, semileptonic decay, NP effects

\end{abstract}

\maketitle

\section{Introduction}

The $B_{(s)}$ to charmed meson semileptonic decays are important for measurements of the CKM matrix element $|V_{cb}|$ and are also probes for new physics (NP) beyond the standard model (SM). Despite being a charged current channel, some intriguing hints of discrepancies have been observed by several experimental collaborations. Measurements of the ratios of branching fractions,
\begin{eqnarray}\label{RDDstar}
R(D^{(*)})= \frac{Br(B \rightarrow D^{(*)} \tau \bar{\nu}_\tau ) }{Br(B \rightarrow D^{(*)} l \bar{\nu}_l ) }, \hspace{1cm} l=e,\mu
\end{eqnarray}
show a $3.3 \sigma$ tension with their SM expectations when the $D$ and $D^*$ results are combined\cite{HFLAV}, which may imply violation of lepton flavor universality. To further confirm or rule out these hints, it is necessary to investigate the additional decay modes mediated by the same parton level transition, not only because these decays can give complementary information, but also because they constitute important backgrounds to $R(D^{(*)})$ measurements. Moreover, better theoretical control of these modes will help improve the determinations of $|V_{cb}|$ and understand the composition of inclusive $B_{(s)} \rightarrow X_c l \bar{\nu}_l$ decays in terms of relevant exclusive channels.

In this study, we focus on the $B_{(s)} \rightarrow D^{**}_{(s)} l \bar{\nu}_l$ decays, with $D^{**}_{(s)}$ denoting P-wave excited charmed mesons. Specifically\cite{PDG2022},
\begin{eqnarray}
& & D^{**} \in \{ D^*_0(2300), D_1(2430), D_1(2420), D^*_2(2460) \}, \label{D**}\\
& & D^{**}_s \in \{ D^*_{s0}(2317), D_{s1}(2460), D_{s1}(2536), D^*_{s2}(2573) \}.  \label{Ds**}
\end{eqnarray}

In the quark model, these mesons can be viewed as constituent quark-antiquark pairs with a total orbital angular momentum $L=1$. (Note that the structures of $D^*_{s0}(2317)$ and $D_{s1}(2460)$ are not completely clear, and we simply interpret them as the lightest orbitally excited states of quark-antiquark pairs for consistency here.) For a hadron containing a single heavy quark, the heavy quark is approximately decoupled. Therefore, the above excited charmed mesons can be classified by the total momentum and parity of the light degrees of freedom $j^P_l$. The first and last two mesons for $D^{**}_{(s)}$ (c.f. Eqs.(\ref{D**}), (\ref{Ds**})) have
$j^P_l = \frac{1}{2}^+$ and $j^P_l = \frac{3}{2}^+$ respectively, which are denoted as $D^{1/2+}_{(s)}$ and $D^{3/2+}_{(s)}$, respectively, in the following.
There is a long-standing interesting `$1/2$ vs $3/2$ puzzle' that theoretical predictions for the branching fractions of semileptonic B decays
into $D^{1/2+}$ are considerably smaller than those into $D^{3/2+}$, i.e. $Br ( B \rightarrow D^{1/2+} l \bar{\nu}_l) \ll Br ( B \rightarrow D^{3/2+} l \bar{\nu}_l)$, conflicting with the experimental results, $Br ( B \rightarrow D^{1/2+} l \bar{\nu}_l) \approx Br ( B \rightarrow D^{3/2+} l \bar{\nu}_l)$\cite{puzzle1,puzzle2,HQETE5}. Our studies may help understand this puzzle.

For exclusive semileptonic decays, the non-perturbative contributions can be parameterized in terms of form factors. The $B_{(s)} \rightarrow D^{**}_{(s)}$ form factors were initially estimated in the Isgur-Scora-Grinstein-Wise (ISGW) quark model and its improved version ISGW2\cite{ISGW,ISGW21, ISGW22}. They have also been calculated via the covariant light-front quark model (LFQM)\cite{LFQM1, LFQM2}.
Some model independent predictions for these decays can be obtained based on heavy quark symmetry.
In Ref.\cite{HQETE1,HQETE2,HQETE3,HQETE4,HQETE5}, with the available experimental results as inputs, the semileptonic $B_{(s)}$ decays into excited charmed mesons and relevant form factors were investigated in the usual heavy quark effective theory (HQET), including the next leading order corrections of heavy quark expansion and NP effects. The form factors of $B \rightarrow D^{**}$ decays were also studied using QCD sum rules in HQET\cite{HQETSR1,HQETSR2}. Additionally, the $B \rightarrow D_1(2430), D_1(2420), D^*_2(2460)$ form factors were evaluated via light cone sum rules (LCSR) and applied to the analysis of relevant semileptonic decays\cite{LCSRB1,LCSRB2,LCSRB3}.

Because heavy quark-antiquark coupling effects in the finite mass corrections are not considered in HQET\cite{HQEFT1,HQEFT2,HQEFT3}, the $B \rightarrow D^{**}$ form factors were calculated to the next leading order of heavy quark expansion in heavy quark effective field theory (HQEFT) with QCD sum rules\cite{HQEFTSR1,HQEFTSR2}.
In HQEFT, all the odd powers of the transverse momentum operator ${D \hspace{-0.3cm} \slash}_\perp$ in the effective current are absent, and thus the forms of the operators become similar to those in the effective Lagrangian. For this reason, fewer universal wave functions are involved. In this study, we intend to give a systematic calculation for the $B_{(s)} \rightarrow D^{**}_{(s)}$ form factors using QCD sum rules in HQEFT and perform a model independent analysis of relevant semileptonic decays, including the contributions from possible NP effects.

The remainder of this paper is organized as follows. In Section
II, we give the definitions of form factors and derive
their expressions in terms of several universal wave functions using
heavy quark expansion to the next leading order in HQEFT. These universal functions can be evaluated via QCD sum rules. The numerical results and discussions of
the form factors are presented in Section III. Based on these form
factors, we predict the differential decay widths, branching fractions, and ratios of branching fractions $R(D^{**}_{(s)})$ for all the relevant semileptonic decays in
Section IV. Section V presents our summary.

\section{Definitions of form factors and formulation via heavy quark expansion and QCD sum rules in HQEFT}

\subsection{Definitions of form factors}

As in Ref.\cite{HQETE4}, we consider the $B_{(s)} \rightarrow D^{**}_{(s)}$ matrix elements of operators with all possible Dirac structures, i.e.
\begin{eqnarray}
O_V = \bar{c} \gamma^\mu b, \hspace{0.5cm} O_A =\bar{c} \gamma^\mu \gamma^5 b, \hspace{0.5cm} O_S = \bar{c} b, \hspace{0.5cm}
O_P = \bar{c} \gamma^5 b, \hspace{0.5cm} O_T = \bar{c} \sigma^{\mu \nu} b,
\end{eqnarray}
 where $\sigma^{\mu \nu} = \frac{i}{2} \left [ \gamma^\mu, \gamma^\nu \right ]$ and $\sigma^{\mu \nu} \gamma^5 = \frac{i}{2} \epsilon^{\mu \nu \rho \sigma} \sigma_{\rho \sigma}$. In the following, for simplicity, we denote $D^*_0(2300), D_1(2430), D_1(2420), D^*_2(2460)$ as $D^*_0, D^\prime_1, D_1, D^*_2$, respectively, and similarly for their strange counterparts. The hadronic matrix elements of these operators can be parameterized in terms of form factors.
For the $B_{(s)} \rightarrow D^{1/2+}_{(s)}$ decays,
\begin{eqnarray}
& & \langle D^*_{(s)0} (v^\prime) | \bar{c} \gamma^5 b | B_{(s)} (v) \rangle = \sqrt{m_{D^*_0} m_B} g_P,  \\
& & \langle D^*_{(s)0} (v^\prime) | \bar{c} \gamma^\mu \gamma^5 b | B_{(s)} (v) \rangle = \sqrt{m_{D^*_0} m_B} [ g_+ ( v^\mu + v^{\prime \mu}) + g_- ( v^\mu - v^{\prime \mu}) ],\\
& & \langle D^*_{(s)0} (v^\prime) | \bar{c} \sigma^{\mu \nu} b | B_{(s)} (v) \rangle =  \sqrt{m_{D^*_0} m_B}  g_T \varepsilon^{\mu \nu \alpha \beta}
v_\alpha v^\prime_\beta, \\
& & \langle D^\prime_{(s)1} ( v^\prime, \epsilon^* ) | \bar{c} b | B_{(s)} (v) \rangle = - \sqrt{m_{D^*_1} m_B} g_S ( \epsilon^* \cdot v ),\\
& & \langle D^\prime_{(s)1} ( v^\prime, \epsilon^* ) | \bar{c} \gamma^\mu b | B_{(s)} (v) \rangle = \sqrt{m_{D^*_1} m_B} [ g_{V_1} \epsilon^{* \mu} +
( g_{V_2} v^\mu + g_{V_3} v^{\prime \mu} ) ( \epsilon^* \cdot v ) ],\\
& & \langle D^\prime_{(s)1} ( v^\prime, \epsilon^* ) | \bar{c} \gamma^\mu \gamma^5 b | B_{(s)} (v) \rangle = i \sqrt{m_{D^*_1} m_B} g_A \varepsilon^{\mu \alpha \beta \gamma} \epsilon^*_\alpha v_\beta v^\prime_\gamma, \\
& & \langle D^*_{(s)1} ( v^\prime, \epsilon^* ) | \bar{c} \sigma^{\mu \nu} b | B_{(s)} (v) \rangle = i \sqrt{m_{D^*_1} m_B}
[ g_{T_1} ( \epsilon^{* \mu} v^\nu - \epsilon^{* \nu } v^\mu ) + g_{T_2} ( \epsilon^{*\mu} v^{\prime \nu} - \epsilon^{* \nu} v^{\prime \mu} )\nonumber \\
& & \hspace{4.5cm} + g_{T_3} ( \epsilon^* \cdot v ) ( v^\mu v^{\prime \nu} - v^\nu v^{\prime \mu} ) ].
\end{eqnarray}
For the $B \rightarrow D^{3/2+}_{(s)}$ decays,
\begin{eqnarray}
& & \langle D_{(s)1} ( v^\prime, \epsilon^* ) | \bar{c} b | B_{(s)} (v) \rangle = \sqrt{m_{D_1} m_B } f_S ( \epsilon^* \cdot v ), \\
& & \langle D_{(s)1} ( v^\prime, \epsilon^* ) | \bar{c}\gamma^\mu b | B_{(s)} (v) \rangle = \sqrt{m_{D_1} m_B } [ f_{V_1} \epsilon^{* \mu} +
( f_{V_2} v^\mu + f_{V_3} v^{\prime \mu} ) ( \epsilon^* \cdot v ) ], \\
& & \langle D_{(s)1} ( v^\prime, \epsilon^* ) | \bar{c}\gamma^\mu \gamma^5 b | B_{(s)} (v) \rangle = i \sqrt{m_{D_1} m_B } f_A \varepsilon^{\mu \alpha \beta \gamma } \epsilon^*_\alpha v_\beta v^\prime_\gamma, \\
& & \langle D_{(s)1} ( v^\prime, \epsilon^* ) | \bar{c} \sigma^{\mu \nu} b | B_{(s)} (v) \rangle = i \sqrt{m_{D_1} m_B } [ f_{T_1} ( \epsilon^{* \mu} v^\nu - \epsilon^{* \nu } v^\mu ) + f_{T_2} ( \epsilon^{*\mu} v^{\prime \nu} - \epsilon^{* \nu} v^{\prime \mu} )\nonumber \\
& & \hspace{4.5cm} + f_{T_3} ( \epsilon^* \cdot v ) ( v^\mu v^{\prime \nu} - v^\nu v^{\prime \mu} ) ],\\
& & \langle D^*_{(s)2} ( v^\prime, \epsilon^* ) | \bar{c} \gamma^5 b| B_{(s)} ( v) \rangle = \sqrt{m_{D^*_2} m_B } k_P \epsilon^*_{\alpha \beta} v^\alpha v^\beta, \\
& & \langle D^*_{(s)2} ( v^\prime, \epsilon^* ) | \bar{c} \gamma^\mu b| B_{(s)} ( v) \rangle = i \sqrt{m_{D^*_2} m_B } k_V \varepsilon^{\mu \alpha \beta
\gamma} \epsilon^*_{\alpha \sigma} v^\sigma v_\beta v^\prime_\gamma, \\
& & \langle D^*_{(s)2} ( v^\prime, \epsilon^* ) | \bar{c} \gamma^\mu \gamma^5 b| B_{(s)} ( v) \rangle = \sqrt{m_{D^*_2} m_B } [ k_{A_1}
\epsilon^{* \mu \alpha} v_\alpha + ( k_{A_2} v^\mu + k_{A_3} v^{\prime \mu} ) \epsilon^*_{\alpha \beta} v^\alpha v^\beta ], \\
& & \langle D^*_{(s)2} ( v^\prime, \epsilon^* ) | \bar{c} \sigma^{\mu \nu} b| B_{(s)} ( v) \rangle = \sqrt{m_{D^*_2} m_B } \varepsilon^{\mu \nu \alpha \beta} \{ [ k_{T_1} ( v+ v^\prime )_\alpha + k_{T_2} ( v - v^\prime)_\alpha ] \epsilon^*_{ \beta \gamma} v^\gamma \nonumber \\
& & \hspace{4.5cm} + k_{T_3} v_\alpha v^\prime_\beta \epsilon^*_{\rho \sigma} v^\rho v^\sigma \}.
\end{eqnarray}

The scalar form factors $g_P$, $g_S$, $f_S$, $k_P$ and tensor form factors $g_T$, $g_{T_i}$, $f_{T_i}$, $k_{T_i}$ $(i=1, 2, 3)$ are only relevant for possible NP effects.

\subsection{Formulation via heavy quark expansion and QCD sum rules in HQEFT}

Now, let us derive the expressions for the form factors using heavy quark expansion and QCD sum rules in HQEFT following similar procedures detailed in Ref.\cite{HQEFTSR1,HQEFTSR2}. The hadronic matrix elements can be expanded over the inverse of heavy quark mass, i.e. $1/m_Q$.
To the next leading order,
\begin{eqnarray}
& & \langle D^{**}_{(s)}(v^\prime) | \bar{c} \Gamma b | B_{(s)}(v)\rangle = \sqrt{\frac{m_{D^{**}_{(s)}} m_{B_{(s)}}}{\bar{\Lambda}_{D^{**}_{(s)}} \bar{\Lambda}_{B_{(s)}}}} \left [
\langle H^\prime_{v^\prime} | \bar{Q}^+_{v^\prime} \Gamma Q^+_v | H_v \rangle \right. \nonumber \\
& & \hspace{0.5cm} + \frac{1}{2m_b} \left ( \langle H^\prime_{v^\prime} | \bar{Q}^+_{v^\prime} \Gamma \frac{P_+}{i v \cdot D} D^2_\perp Q^+_v | H_v \rangle + \langle H^\prime_{v^\prime} | \bar{Q}^+_{v^\prime} \Gamma \frac{P_+}{i v \cdot D} \frac{i}{2} \sigma_{\alpha \beta} F^{\alpha \beta} Q^+_v | H_v \rangle \right ) \nonumber \\
& & \hspace{0.5cm} \left. +  \frac{1}{2m_c} \left ( \langle H^\prime_{v^\prime} | \bar{Q}^+_{v^\prime} \overleftarrow{D}^2_\perp \frac{P^\prime_+}{ -i v^\prime \cdot \overleftarrow{D}} \Gamma Q^+_v | H_v \rangle + \langle H^\prime_{v^\prime} | \bar{Q}^+_{v^\prime} \frac{i}{2} \sigma_{\alpha \beta} F^{\alpha \beta} \frac{P^\prime_+}{ -i v^\prime \cdot \overleftarrow{D}} \Gamma Q^+_v | H_v \rangle \right )\right ], \nonumber \\
\end{eqnarray}
where $\Gamma$ is an arbitrary combination of Dirac matrices, and $P^{(\prime)}_+ = (1+
\not\!{v}^{(\prime)})/2$.

The relevant matrix elements in HQEFT can be represented by a set of universal functions. For the $B_{(s)} \rightarrow D^{1/2+}_{(s)}$ decays,
\begin{eqnarray}
& & \langle H^\prime_{v^\prime} | \bar{Q}^+_{v^\prime} \Gamma Q^+_v | H_v \rangle = \zeta  Tr\left[ \bar{{\cal K}}_{v^\prime} \Gamma {\cal M}_v \right ], \\
& & \langle H^\prime_{v^\prime} | \bar{Q}^+_{v^\prime} \Gamma \frac{P_+}{i v \cdot D} D^2_\perp Q^+_v | H_v \rangle =
-   \frac{\chi^b_0}{\bar{\Lambda}_{(s)}} Tr \left [ \bar{{\cal K}}_{v^\prime} \Gamma {\cal M}_v \right ], \\
& & \langle H^\prime_{v^\prime} | \bar{Q}^+_{v^\prime} \overleftarrow{D}^2_\perp \frac{P^\prime_+}{ -i v^\prime \cdot \overleftarrow{D}} \Gamma Q^+_v | H_v \rangle = -   \frac{\chi^c_0}{\bar{\Lambda}^{1/2}_{(s)}} Tr \left [ \bar{{\cal K}}_{v^\prime} \Gamma {\cal M}_v \right ],\\
& & \langle H^\prime_{v^\prime} | \bar{Q}^+_{v^\prime} \Gamma \frac{P_+}{i v \cdot D} \frac{i}{2} \sigma_{\alpha \beta} F^{\alpha \beta} Q^+_v | H_v \rangle= - \frac{1}{\bar{\Lambda}_{(s)}} Tr \left [ R^b_{\alpha \beta} (v, v^\prime) \bar{{\cal K}}_{v^\prime} \Gamma P_+ i \sigma^{\alpha \beta} {\cal M}_v \right ], \\
& & \langle H^\prime_{v^\prime} | \bar{Q}^+_{v^\prime} \frac{i}{2} \sigma_{\alpha \beta} F^{\alpha \beta} \frac{P^\prime_+}{ -i v^\prime \cdot \overleftarrow{D}} \Gamma Q^+_v | H_v \rangle = - \frac{1}{\bar{\Lambda}^{1/2}_{(s)}} Tr \left[ R^c_{\alpha \beta} (v, v^\prime) \bar{{\cal K}}_{v^\prime} i \sigma^{\alpha \beta} P^\prime_+ \Gamma {\cal M}_v \right ],
\end{eqnarray}
where
\begin{eqnarray}
& & R^b_{\alpha \beta}(v, v^\prime) = \chi^b_1  \gamma_\alpha \gamma_\beta + \chi^b_2  v^\prime_\alpha \gamma_\beta, \\
& & R^c_{\alpha \beta}(v, v^\prime) = \chi^c_1  \gamma_\alpha \gamma_\beta + \chi^c_2  v_\alpha \gamma_\beta.
\end{eqnarray}
Similarly, for the $B_{(s)} \rightarrow D^{3/2+}_{(s)}$ decays,
\begin{eqnarray}
& & \langle H^\prime_{v^\prime} | \bar{Q}^+_{v^\prime} \Gamma Q^+_v | H_v \rangle = \tau Tr\left[ v_\sigma \bar{{\cal F}}^\sigma_{v^\prime} \Gamma {\cal M}_v \right ], \\
& & \langle H^\prime_{v^\prime} | \bar{Q}^+_{v^\prime} \Gamma \frac{P_+}{i v \cdot D} D^2_\perp Q^+_v | H_v \rangle =
-  \frac{\eta^b_0 }{\bar{\Lambda}_{(s)}} Tr \left [ v_\sigma \bar{{\cal F}}^\sigma_{v^\prime} \Gamma {\cal M}_v \right ], \\
& & \langle H^\prime_{v^\prime} | \bar{Q}^+_{v^\prime} \overleftarrow{D}^2_\perp \frac{P^\prime_+}{ -i v^\prime \cdot \overleftarrow{D}} \Gamma Q^+_v | H_v \rangle = -  \frac{\eta^c_0}{\bar{\Lambda}^{3/2}_{(s)}} Tr \left [ v_\sigma \bar{{\cal F}}^\sigma_{v^\prime} \Gamma {\cal M}_v \right ],\\
& & \langle H^\prime_{v^\prime} | \bar{Q}^+_{v^\prime} \Gamma \frac{P_+}{i v \cdot D} \frac{i}{2} \sigma_{\alpha \beta} F^{\alpha \beta} Q^+_v | H_v \rangle= - \frac{1}{\bar{\Lambda}_{(s)}} Tr \left [ R^b_{\sigma \alpha \beta} (v, v^\prime) \bar{{\cal F}}^\sigma_{v^\prime} \Gamma P_+ i \sigma^{\alpha \beta} {\cal M}_v \right ], \\
& & \langle H^\prime_{v^\prime} | \bar{Q}^+_{v^\prime} \frac{i}{2} \sigma_{\alpha \beta} F^{\alpha \beta} \frac{P^\prime_+}{ -i v^\prime \cdot \overleftarrow{D}} \Gamma Q^+_v | H_v \rangle = - \frac{1}{\bar{\Lambda}^{3/2}_{(s)}} Tr \left[ R^c_{\sigma \alpha \beta} (v, v^\prime) \bar{{\cal F}}^\sigma_{v^\prime} i \sigma^{\alpha \beta} P^\prime_+ \Gamma {\cal M}_v \right ],
\end{eqnarray}
where
\begin{eqnarray}
& & R^b_{\sigma \alpha \beta} (v, v^\prime) = \eta^b_1  v_\sigma \gamma_\alpha \gamma_\beta + \eta^b_2 v_\sigma v^\prime_\alpha \gamma_\beta + \eta^b_3 g_{\sigma \alpha} v^\prime_\beta, \\
& & R^c_{\sigma \alpha \beta} (v, v^\prime) = \eta^c_1  v_\sigma \gamma_\alpha \gamma_\beta + \eta^c_2  v_\sigma v_\alpha \gamma_\beta + \eta^c_3 g_{\sigma \alpha} v_\beta.
\end{eqnarray}

The universal wave functions $\zeta$, $\tau$, $\chi^{b(c)}_i (i =0, 1, 2)$, and $\eta^{b(c)}_j (j =0, 1, 2, 3)$ depend on $\omega = v \cdot v^\prime = ( m^2_{B_{(s)}} + m^{** 2}_{D_{(s)}} - q^2 )/(2 m_{B_{(s)}} m_{D^{**}_{(s)}})$, and the spin wave functions for the initial and final state mesons ${\cal M}_v$, ${\cal K}_{v^\prime}$ and ${\cal F}^\mu_{v^\prime}$ have the following
forms:
\begin{eqnarray}
& & {\cal M}_v = - \sqrt{\bar{\Lambda}_{(s)}} P_+ \gamma^5, \hspace{0.5cm} \mbox{for $B_{(s)}$}, \\
& & {\cal K}_{v^\prime} = \sqrt{\bar{\Lambda}^{1/2}_{(s)}} P^\prime_+  \left\{
       \begin{array}{cl}
       1, & \mbox{for $D^*_{(s)0}$}  \\
      -\epsilon\hspace{-0.15cm}\slash \gamma^5, & \mbox{for $D^\prime_{(s)1}$}
           \end{array}
         \right. ,\\
& & {\cal F}^\mu_{v^\prime} = \sqrt{\bar{\Lambda}^{3/2}_{(s)}} P^\prime_+ \left\{
       \begin{array}{cl}
       - \sqrt{\frac{3}{2}} \gamma^5 \epsilon^\nu \left[ g^\mu_\nu - \frac{1}{3} \gamma_\nu \left ( \gamma^\mu - v^{\prime \mu} \right ) \right ], & \mbox{for $D_{(s)1}$}  \\
      \epsilon^{\mu \nu} \gamma_\nu, & \mbox{for $D^*_{(s)2}$}
           \end{array}
         \right.,
\end{eqnarray}
 where $\bar{{\cal K}}_{v^\prime} = \gamma^0 {\cal K}_{v^\prime}^\dagger \gamma^0$ and $\bar{{\cal F}}^\mu_{v^\prime} = \gamma^0 {\cal F}^{\mu \dagger}_{v^\prime} \gamma^0$. $\bar{\Lambda}_{(s)}, \bar{\Lambda}^{1/2}_{(s)}, \bar{\Lambda}^{3/2}_{(s)}$ are the heavy flavor independent binding energies of $j^P_l= \frac{1}{2}^-, \frac{1}{2}^+, \frac{3}{2}^+$ heavy mesons, respectively.

Then, the form factors can be expressed in terms of the universal wave functions up to the next leading order of heavy quark expansion. For the $B_{(s)} \rightarrow D^{1/2+}_{(s)}$ decays,
\begin{eqnarray}
& & g_P = -(1-\omega)  \left [ \tilde{\zeta} + \frac{\zeta}{2 m_b\bar{\Lambda}_{(s)}}  ( \kappa_1 (1) + 3 \kappa_2 (1)  ) + \frac{\zeta}{2  m_c \bar{\Lambda}^{1/2}_{(s)} }  ( \kappa^{1/2}_1 (1) + 3 \kappa^{1/2}_2 (1)  ) \right. \nonumber \\
& & \hspace{1cm} \left. -\frac{1}{ m_b \bar{\Lambda}_{(s)}} \chi^b - \frac{1}{ m_c \bar{\Lambda}^{1/2}_{(s)}} \left ( 3  \chi^c_1 - (1+\omega) \chi^c_2 \right ) \right ], \label{gP}\\
& & g_+ =0, \label{g+}\\
& & g_- = g_T=  \tilde{\zeta}   + \frac{\zeta}{2 m_b\bar{\Lambda}_{(s)}}  ( \kappa_1 (1) + 3 \kappa_2 (1)  )  + \frac{\zeta}{2  m_c \bar{\Lambda}^{1/2}_{(s)} }  ( \kappa^{1/2}_1 (1) + 3 \kappa^{1/2}_2 (1)  )\nonumber \\
& & \hspace{1cm}- \frac{1}{m_b \bar{\Lambda}_{(s)}}  \chi^b  -  \frac{1}{ m_c \bar{\Lambda}^{1/2}_{(s)}}\left [ 3 \chi^c_1 - (1+\omega) \chi^c_2 \right ],\\
& & g_S=  \tilde{\zeta} + \frac{\zeta}{2 m_b\bar{\Lambda}_{(s)}}  ( \kappa_1 (1) + 3 \kappa_2 (1)  )  + \frac{\zeta}{2  m_c \bar{\Lambda}^{1/2}_{(s)} }  ( \kappa^{1/2}_1 (1) -  \kappa^{1/2}_2 (1)  )-  \frac{1}{m_b \bar{\Lambda}_{(s)}} \chi^b \nonumber \\
& & \hspace{1cm}  + \frac{1}{ m_c \bar{\Lambda}^{1/2}_{(s)}}\left [ \chi^c_1 - (1+\omega) \chi^c_2 \right ], \\
& & g_{V_1}=  -(1- \omega) \left [ \tilde{\zeta} + \frac{\zeta}{2 m_b\bar{\Lambda}_{(s)}}  ( \kappa_1 (1) + 3 \kappa_2 (1)  )  + \frac{\zeta}{2  m_c \bar{\Lambda}^{1/2}_{(s)} }  ( \kappa^{1/2}_1 (1) -  \kappa^{1/2}_2 (1)  ) \right. \nonumber \\
& & \left. \hspace{1cm} - \frac{1}{ m_b \bar{\Lambda}_{(s)}} \chi^b + \frac{1}{ m_c \bar{\Lambda}^{1/2}_{(s)}} \chi^c_1 \right ],   \\
& & g_{V_2}=-g_{T_3}= \frac{\chi^c_2}{ m_c \bar{\Lambda}^{1/2}_{(s)}},  \\
& & g_{V_3}= -\tilde{\zeta} - \frac{\zeta}{2 m_b\bar{\Lambda}_{(s)}}  ( \kappa_1 (1) + 3 \kappa_2 (1)  )  - \frac{\zeta}{2  m_c \bar{\Lambda}^{1/2}_{(s)} }  ( \kappa^{1/2}_1 (1) -  \kappa^{1/2}_2 (1)  )  \nonumber \\
& & \hspace{1cm} + \frac{1}{ m_b \bar{\Lambda}_{(s)}} \chi^b - \frac{1}{ m_c \bar{\Lambda}^{1/2}_{(s)}}( \chi^c_1 - \chi^c_2 ), \\
& & g_A = -g_{T_1}= g_{T_2}=\tilde{\zeta}  + \frac{\zeta}{2 m_b\bar{\Lambda}_{(s)}}  ( \kappa_1 (1) +  3 \kappa_2 (1)  )+ \frac{\zeta}{2  m_c \bar{\Lambda}^{1/2}_{(s)} }  ( \kappa^{1/2}_1 (1) -  \kappa^{1/2}_2 (1)  )\nonumber \\
& & \hspace{1cm} - \frac{1}{ m_b \bar{\Lambda}_{(s)}}\chi^b + \frac{1}{ m_c \bar{\Lambda}^{1/2}_{(s)}} \chi^c_1,
\end{eqnarray}
where
\begin{eqnarray}
& & \tilde{\zeta} = \zeta -  \frac{\chi^b_0}{2 m_b \bar{\Lambda}_{(s)}} - \frac{\chi^c_0}{2 m_c \bar{\Lambda}^{1/2}_{(s)}}, \\
& & \chi^b =  3 \chi^b_1 - (1+\omega) \chi^b_2.
\end{eqnarray}
For the $B_{(s)} \rightarrow D^{3/2+}_{(s)}$ decays,
\begin{eqnarray}
& & f_S = - \frac{2}{\sqrt{6}} (1+\omega)  \left [ \tilde{\tau}  + \frac{\tau}{2 m_b\bar{\Lambda}_{(s)}}  ( \kappa_1 (1) +  3 \kappa_2 (1)  )+ \frac{\tau}{2  m_c \bar{\Lambda}^{3/2}_{(s)} }  ( \kappa^{3/2}_1 (1) +5  \kappa^{3/2}_2 (1)  ) \right. \nonumber \\
& & \hspace{1cm} \left.+ \frac{1}{ m_b \bar{\Lambda}_{(s)}} \eta^b  + \frac{1}{ m_c \bar{\Lambda}^{3/2}_{(s)}} \left ( -3 \eta^c_1 + (1-\omega) \eta^c_2  + \frac{1}{2} \eta^c_3 \right ) \right ], \label{fs}\\
& & f_{V_1} =  \frac{1}{\sqrt{6}} (1-\omega^2)  \left [ \tilde{\tau}  + \frac{\tau}{2 m_b\bar{\Lambda}_{(s)}}  ( \kappa_1 (1) +  3 \kappa_2 (1)  )+ \frac{\tau}{2  m_c \bar{\Lambda}^{3/2}_{(s)} }  ( \kappa^{3/2}_1 (1) +5  \kappa^{3/2}_2 (1)  ) \right. \nonumber \\
& & \hspace{1cm} \left. + \frac{1}{ m_b \bar{\Lambda}_{(s)}}  \eta^b + \frac{1}{ m_c \bar{\Lambda}^{3/2}_{(s)}}( \eta^c_1 + \frac{3}{2} \eta^c_3 ) \right ],  \\
& & f_{V_2} = - \frac{3}{\sqrt{6}}\left [ \tilde{\tau}  + \frac{\tau}{2 m_b\bar{\Lambda}_{(s)}}  ( \kappa_1 (1) +  3 \kappa_2 (1)  )+ \frac{\tau}{2  m_c \bar{\Lambda}^{3/2}_{(s)} }  ( \kappa^{3/2}_1 (1) +5  \kappa^{3/2}_2 (1)  )\right ] \nonumber \\
& & \hspace{1cm} - \frac{1}{\sqrt{6}} \frac{3}{ m_b \bar{\Lambda}_{(s)}}  \eta^b - \frac{1}{\sqrt{6}}\frac{5}{ m_c \bar{\Lambda}^{3/2}_{(s)}} \left [ - \eta^c_1 +\frac{2}{5} ( 1-\omega) \eta^c_2 + \frac{1}{2} \eta^c_3 \right ],\\
& & f_{V_3} = \frac{1}{\sqrt{6}} (\omega-2)  \left [ \tilde{\tau}  + \frac{\tau}{2 m_b\bar{\Lambda}_{(s)}}  ( \kappa_1 (1) +  3 \kappa_2 (1)  )+ \frac{\tau}{2  m_c \bar{\Lambda}^{3/2}_{(s)} }  ( \kappa^{3/2}_1 (1) +5  \kappa^{3/2}_2 (1)  )  \right ] \nonumber \\
& & \hspace{1cm} + \frac{1}{\sqrt{6}} \frac{1}{ m_b \bar{\Lambda}_{(s)}}  (\omega-2) \eta^b +  \frac{1}{\sqrt{6}} \frac{1}{ m_c \bar{\Lambda}^{3/2}_{(s)}} \left [ (6+\omega) \eta^c_1 - 2 ( 1-\omega) \eta^c_2 - ( 1- \frac{3}{2} \omega ) \eta^c_3 \right ],\\
& & f_A = -f_{T_1} = f_{T_2} =  - \frac{1}{\sqrt{6}} (1+\omega)\left [ \tilde{\tau}  + \frac{\tau}{2 m_b\bar{\Lambda}_{(s)}}  ( \kappa_1 (1) +  3 \kappa_2 (1)  )\right. \nonumber \\
& & \hspace{1cm} \left.+ \frac{\tau}{2  m_c \bar{\Lambda}^{3/2}_{(s)} }  ( \kappa^{3/2}_1 (1) +5  \kappa^{3/2}_2 (1)  )  + \frac{1}{ m_b \bar{\Lambda}_{(s)}} \eta^b + \frac{1}{ m_c \bar{\Lambda}^{3/2}_{(s)}} ( \eta^c_1 + \frac{3}{2} \eta^c_3  ) \right ],\\
& & f_{T_3} = \frac{3}{\sqrt{6}} \left [ \tilde{\tau}  + \frac{\tau}{2 m_b\bar{\Lambda}_{(s)}}  ( \kappa_1 (1) +  3 \kappa_2 (1)  )+ \frac{\tau}{2  m_c \bar{\Lambda}^{3/2}_{(s)} }  ( \kappa^{3/2}_1 (1) +5  \kappa^{3/2}_2 (1)  ) \right. \nonumber \\
& & \hspace{1cm} \left.+ \frac{1}{ m_b \bar{\Lambda}_{(s)}} \eta^b - \frac{1}{6} \frac{1}{ m_c \bar{\Lambda}^{3/2}_{(s)}} [ 10 \eta^c_1 - 4 (1-\omega) \eta^c_2 - 5 \eta^c_3 ]\right ], \\
& & k_P = \tilde{\tau}  + \frac{\tau}{2 m_b\bar{\Lambda}_{(s)}}  ( \kappa_1 (1) +  3 \kappa_2 (1)  )+ \frac{\tau}{2  m_c \bar{\Lambda}^{3/2}_{(s)} }  ( \kappa^{3/2}_1 (1) -  3 \kappa^{3/2}_2 (1)  ) + \frac{1}{ m_b \bar{\Lambda}_{(s)}} \eta^b \nonumber \\
& & \hspace{1cm} + \frac{1}{ m_c \bar{\Lambda}^{3/2}_{(s)}} \left [ \eta^c_1 - (1-\omega) \eta^c_2 - \frac{1}{2} \eta^c_3 \right ],\\
& & k_V = - \left [ \tilde{\tau}  + \frac{\tau}{2 m_b\bar{\Lambda}_{(s)}}  ( \kappa_1 (1) +  3 \kappa_2 (1)  )+ \frac{\tau}{2  m_c \bar{\Lambda}^{3/2}_{(s)} }  ( \kappa^{3/2}_1 (1) -  3 \kappa^{3/2}_2 (1)  ) \right ]\nonumber \\
& & \hspace{1cm} - \frac{1}{ m_b \bar{\Lambda}_{(s)}}\eta^b  - \frac{1}{ m_c \bar{\Lambda}^{3/2}_{(s)}}\left (  \eta^c_1 - \frac{1}{2} \eta^c_3 \right ), \\
& & k_{A_1} = - (1+\omega) \left [ \tilde{\tau}  + \frac{\tau}{2 m_b\bar{\Lambda}_{(s)}}  ( \kappa_1 (1) +  3 \kappa_2 (1)  )+ \frac{\tau}{2  m_c \bar{\Lambda}^{3/2}_{(s)} }  ( \kappa^{3/2}_1 (1) -  3 \kappa^{3/2}_2 (1)  ) \right. \nonumber \\
& & \hspace{1cm} \left. +  \frac{1}{ m_b \bar{\Lambda}_{(s)}} \eta^b + \frac{1}{ m_c \bar{\Lambda}^{3/2}_{(s)}}\left (   \eta^c_1 -\frac{1}{2} \eta^c_3\right )\right ],  \\
& & k_{A_2} = k_{T_3} =\frac{1}{ m_c \bar{\Lambda}^{3/2}_{(s)}} \eta^c_2, \\
& & k_{A_3} =  \tilde{\tau}  + \frac{\tau}{2 m_b\bar{\Lambda}_{(s)}}  ( \kappa_1 (1) +  3 \kappa_2 (1)  )+ \frac{\tau}{2  m_c \bar{\Lambda}^{3/2}_{(s)} }  ( \kappa^{3/2}_1 (1) -  3 \kappa^{3/2}_2 (1)  )\nonumber \\
& & \hspace{1cm}  +  \frac{1}{ m_b \bar{\Lambda}_{(s)}} \eta^b + \frac{1}{ m_c \bar{\Lambda}^{3/2}_{(s)}} \left (  \eta^c_1 - \eta^c_2 - \frac{1}{2} \eta^c_3 \right ), \\
& & k_{T_1}= \tilde{\tau}  + \frac{\tau}{2 m_b\bar{\Lambda}_{(s)}}  ( \kappa_1 (1) +  3 \kappa_2 (1)  )+ \frac{\tau}{2  m_c \bar{\Lambda}^{3/2}_{(s)} }  ( \kappa^{3/2}_1 (1) -  3 \kappa^{3/2}_2 (1)  ) \nonumber \\
& & \hspace{1cm} + \frac{1}{ m_c \bar{\Lambda}^{3/2}_{(s)}}\left (   \eta^c_1 -\frac{1}{2} \eta^c_3\right ),  \\
& & k_{T_2} = \frac{1}{m_b \bar{\Lambda}_{(s)}} \eta^b, \label{kT2}
\end{eqnarray}
where
\begin{eqnarray}
& & \tilde{\tau} = \tau -  \frac{\eta^b_0}{2 m_b \bar{\Lambda}_{(s)}} - \frac{\eta^c_0}{2 m_c \bar{\Lambda}^{3/2}_{(s)}}, \\
& & \eta^b =  -3 \eta^b_1 - ( 1-\omega) \eta^b_2- \frac{1}{2} \eta^b_3. \label{etab}
\end{eqnarray}
For the form factors in the SM, i.e. $g_+$, $g_-$, $g_{V_i}$, $g_A$, $f_{V_i}$, $f_A$, $k_{A_i}$, $k_A$ $(i=1,2,3)$, the expressions agree with those in Ref.\cite{HQEFTSR1,HQEFTSR2}.
The values of $\kappa_1(1), \kappa_2(1), \kappa^{1/2}_1(1), \kappa^{1/2}_2(1), \kappa^{3/2}_1(1), \kappa^{3/2}_2(1)$ and $\bar{\Lambda}_{(s)}$, $\bar{\Lambda}^{1/2}_{(s)}$, $\bar{\Lambda}^{3/2}_{(s)}$ can be extracted by fitting the meson masses. As detailed in APPENDIX~\ref{kappaPs},
\begin{eqnarray}
& & \kappa_1(1)= \frac{m_b m_c}{m_b - m_c} ( \bar{m}_{B_{(s)}} - \bar{m}_{D_{(s)}} -m_b +m_c), \\
& & \kappa_2(1) = \frac{1}{4}m_c ( m_{D^*_{(s)}}-m_{D_{(s)}}), \\
& & \kappa^{1/2}_1(1) = \frac{m_b m_c}{m_b - m_c} ( \bar{m}_{B^{1/2}_{(s)}} - \bar{m}_{D^{1/2}_{(s)}} -m_b +m_c),\\
& & \kappa^{1/2}_2(1) = \frac{1}{4} m_c ( m_{D^\prime_{(s)1}} - m_{D^*_{(s)0}}), \\
& & \kappa^{3/2}_1(1) = \frac{m_b m_c}{m_b - m_c} ( \bar{m}_{B^{3/2}_{(s)}} - \bar{m}_{D^{3/2}_{(s)}} -m_b +m_c),\\
& & \kappa^{3/2}_2(1) = \frac{1}{8} m_c (m_{D^*_{(s)2}}-m_{D_{(s)1}}),
\end{eqnarray}
where the spin average masses of $j^P_l = \frac{1}{2}^-, \frac{1}{2}^+, \frac{3}{2}^+$ doublets
\begin{eqnarray}
& & \bar{m}_{B_{(s)}}= \frac{1}{4}(m_{B_{(s)}} + 3 m_{B^*_{(s)}}), \\
& & \bar{m}_{D_{(s)}}= \frac{1}{4}(m_{D_{(s)}} + 3 m_{D^*_{(s)}}), \\
& & \bar{m}_{B^{1/2}_{(s)}}= \frac{1}{4} ( m_{B^*_{(s)0}} + 3 m_{B^\prime_{(s)1}}), \\
& & \bar{m}_{D^{1/2}_{(s)}}= \frac{1}{4} ( m_{D^*_{(s)0}} + 3 m_{D^\prime_{(s)1}}), \\
& & \bar{m}_{B^{3/2}_{(s)}} = \frac{1}{8}( 3 m_{B_{(s)1}} + 5 m_{B^*_{(s)2}}),\\
& & \bar{m}_{D^{3/2}_{(s)}} = \frac{1}{8}( 3 m_{D_{(s)1}} + 5 m_{D^*_{(s)2}}).
\end{eqnarray}
Additionally, the binding energies
\begin{eqnarray}
& & \bar{\Lambda}_{(s)} = m_{D_{(s)}} - m_c + \frac{1}{m_c} \left ( \kappa_1 (1) + 3 \kappa_2 (1) \right ), \\
& & \bar{\Lambda}^{1/2}_{(s)} = m_{D^*_{(s)0}} - m_c + \frac{1}{m_c} \left ( \kappa^{1/2}_1 (1) + 3 \kappa^{1/2}_2 (1) \right ), \\
& & \bar{\Lambda}^{3/2}_{(s)} = m_{D_{(s)1}} - m_c + \frac{1}{m_c} \left ( \kappa^{3/2}_1 (1) + 5 \kappa^{3/2}_2 (1) \right ).
\end{eqnarray}

As found from Eqs.(\ref{gP})-(\ref{etab}), the form factors simply reduce to the leading order wave functions $\zeta$, $\tau$ in the heavy quark limit. Considering the corrections from the next leading order of heavy quark expansion, 14 more functions $\chi^{b(c)}_i$ $(i=0,1,2)$ and $\eta^{b(c)}_j$ $(j=0,1,2,3)$ are involved. Wave functions with subscript zero and nonzero are defined by kinetic and chromomagnetic operators, respectively.

The universal functions $\zeta$, $\tau$, $\chi^{b(c)}_0$, $\eta^{b(c)}_0$ have been evaluated via QCD sum rules in Ref.\cite{HQEFTSR1,HQEFTSR2}. It is found that
\begin{eqnarray}
& & f_{\frac{1}{2}^+} f_{\frac{1}{2}^-} \zeta e^{- ( \bar{\Lambda}_{(s)} + \bar{\Lambda}^{1/2}_{(s)} )/T} = \frac{1}{8 \pi^2 (1+ \omega)^2} \int^{s^\zeta_0}_0 d \nu \nu^3 e^{-\nu/T} - \frac{2 T }{3 \pi} \alpha_s \langle \bar{q} q \rangle  \nonumber  \\
& & \hspace{2.5cm} + \frac{1}{96 \pi^2 T} \left [ 6 \pi^2 ( \omega +2) - 4 \pi ( \omega +1) \alpha_s \right ] i \langle \bar{q} \sigma_{\alpha \beta} T^a F^{a \alpha \beta} q \rangle \nonumber \\
& & \hspace{2.5cm} + \frac{\omega -1}{ 192 \pi ( \omega +1)} \alpha_s \langle F^a_{\alpha \beta} F^{a \alpha \beta} \rangle \equiv {\cal SR}_\zeta, \label{SRzeta}\\
& & f_{\frac{3}{2}^+} f_{\frac{1}{2}^-} \tau e^{- ( \bar{\Lambda}_{(s)} + \bar{\Lambda}^{3/2}_{(s)} )/T} = \frac{1}{2 \pi^2 ( \omega +1)^3 }  \int^{s^\tau_0}_0 d \nu \nu^3 e^{-\nu/T} + \frac{i}{12 T} \langle \bar{q} \sigma_{\alpha \beta} T^a F^{a \alpha \beta} q \rangle \nonumber \\
& & \hspace{2.5cm} - \frac{\omega +5}{ 96 \pi ( \omega +1)^2} \alpha_s \langle F^a_{\alpha \beta} F^{a \alpha \beta} \rangle \equiv {\cal SR}_\tau, \\
& &  f_{\frac{1}{2}^+} f_{\frac{1}{2}^-} \frac{\chi^b_0}{\bar{\Lambda}_{(s)}} e^{- ( \bar{\Lambda}_{(s)} + \bar{\Lambda}^{1/2}_{(s)} )/T} = - \frac{\omega+4}{16 \pi^2 ( 1+ \omega)^3} \int^{s^b_0}_0 d \nu \nu^4 e^{-\nu/T} \nonumber \\
& & \hspace{2.5cm} - \frac{5 T^2 }{3 \pi ( \omega +1)} \alpha_s \langle \bar{q} q \rangle + \frac{(\omega +2)T}{96 \pi ( 1+ \omega)^2} \alpha_s \langle F^a_{\alpha \beta} F^{a \alpha \beta} \rangle \equiv {\cal SR}_{\chi^b_0}, \\
& &  f_{\frac{1}{2}^+} f_{\frac{1}{2}^-} \frac{\chi^c_0}{\bar{\Lambda}^{1/2}_{(s)}} e^{- ( \bar{\Lambda}_{(s)} + \bar{\Lambda}^{1/2}_{(s)} )/T} = \frac{3 (3 \omega+2)}{16 \pi^2 ( 1+ \omega)^3} \int^{s^c_0}_0 d \nu \nu^4 e^{-\nu/T} \nonumber \\
& & \hspace{2.5cm} - \frac{( 4 \omega +3) T^2 }{3 \pi ( \omega +1)} \alpha_s \langle \bar{q} q \rangle - \frac{(\omega +8)T}{96 \pi ( 1+ \omega)} \alpha_s \langle F^a_{\alpha \beta} F^{a \alpha \beta} \rangle \equiv {\cal SR}_{\chi^c_0}, \\
& &  f_{\frac{3}{2}^+} f_{\frac{1}{2}^-} \frac{\eta^b_0}{\bar{\Lambda}_{(s)}} e^{- ( \bar{\Lambda}_{(s)} + \bar{\Lambda}^{3/2}_{(s)} )/T}= \frac{1+4 \omega}{8 \pi^2 ( 1+ \omega)^4} \int^{s^{ \prime b }_0}_0 d \nu \nu^4 e^{- \nu/T} \nonumber \\
& & \hspace{2.5cm} - \frac{2 T^2}{ 3 \pi ( 1+ \omega)^2} \alpha_s \langle \bar{q} q \rangle - \frac{(7 -\omega)T}{96 \pi ( 1+ \omega)^3 } \alpha_s \langle F^a_{\alpha \beta} F^{a \alpha \beta} \rangle \equiv {\cal SR}_{\eta^b_0}, \\
& &  f_{\frac{3}{2}^+} f_{\frac{1}{2}^-} \frac{\eta^c_0}{\bar{\Lambda}^{3/2}_{(s)}} e^{- ( \bar{\Lambda}_{(s)} + \bar{\Lambda}^{3/2}_{(s)} )/T} = \frac{3 (2+3 \omega)}{8 \pi^2 ( 1+ \omega)^4} \int^{s^{ \prime c }_0}_0 d \nu \nu^4 e^{- \nu/T} \nonumber \\
& & \hspace{2.5cm}  - \frac{2(3 + 2 \omega)T^2}{3 \pi (1+\omega)^2} \alpha_s \langle \bar{q} q \rangle + \frac{(9 \omega +1)T}{96 \pi (1+ \omega)^3} \alpha_s \langle F^a_{\alpha \beta} F^{a \alpha \beta} \rangle \equiv {\cal SR}_{\eta^c_0}.
\end{eqnarray}
For the decay constants $f_{\frac{1}{2}^-}$, $f_{\frac{1}{2}^+}$, and $f_{\frac{3}{2}^+}$,
\begin{eqnarray}
& &  f^2_{\frac{1}{2}^-} e^{-2 \bar{\Lambda}_{(s)}/T} = \frac{3}{16 \pi^2} \int^{s^-_0}_0 d \nu \nu^2 e^{-\nu/T} -
\frac{1}{2} \left ( 1+ \frac{4 \alpha_s}{3 \pi} \right ) \langle \bar{q}q \rangle \nonumber \\
& & \hspace{2.5cm} - \frac{1}{8 T^2} \left ( 1 + \frac{4 \alpha_s}{\pi} \right )
i \langle \bar{q} \sigma_{\alpha \beta} T^a F^{a \alpha \beta} q \rangle - \frac{1}{48 \pi T} \alpha_s \langle F^a_{\alpha \beta} F^{a \alpha \beta} \rangle \equiv {\cal SR}_{\frac{1}{2}^-},    \\
& & f^2_{\frac{1}{2}^+} e^{-2 \bar{\Lambda}^{1/2}_{(s)}/T}= \frac{3}{64 \pi^2} \int^{s^+_0}_0 d \nu \nu^4 e^{-\nu/T} + \left (
\frac{3}{16} - \frac{\alpha_s}{32 \pi} \right ) i \langle \bar{q} \sigma_{\alpha \beta} T^a F^{a \alpha \beta} q \rangle \equiv {\cal SR}_{\frac{1}{2}^+},  \\
& & f^2_{\frac{3}{2}^+} e^{-2 \bar{\Lambda}^{3/2}_{(s)}/T}= \frac{1}{64 \pi^2} \int^{s^{\prime+}_0}_0 d \nu \nu^4 e^{-\nu/T} + \frac{ i}{12} \langle \bar{q} \sigma_{\alpha \beta} T^a F^{a \alpha \beta} q \rangle  \nonumber \\
& & \hspace{2.5cm} - \frac{T}{32 \pi} \alpha_s \langle F^a_{\alpha \beta} F^{a \alpha \beta} \rangle \equiv {\cal SR}_{\frac{3}{2}^+}. \label{SR3/2p}
\end{eqnarray}
From Eqs.(\ref{SRzeta})-(\ref{SR3/2p}), we easily obtain
\begin{eqnarray}
& & \zeta = \frac{{\cal SR}_\zeta}{\sqrt{{\cal SR}_{\frac{1}{2}^-} \times {\cal SR}_{\frac{1}{2}^+}}}, \label{zeta} \\
& & \tau = \frac{{\cal SR}_\tau}{\sqrt{{\cal SR}_{\frac{1}{2}^-} \times {\cal SR}_{\frac{3}{2}^+}}}, \\
& & \frac{\chi^b_0}{\bar{\Lambda}_{(s)}} = \frac{{\cal SR}_{\chi^b_0}}{\sqrt{{\cal SR}_{\frac{1}{2}^-} \times {\cal SR}_{\frac{1}{2}^+}}},\\
& & \frac{\chi^c_0}{\bar{\Lambda}^{1/2}_{(s)}} = \frac{{\cal SR}_{\chi^c_0}}{\sqrt{{\cal SR}_{\frac{1}{2}^-} \times {\cal SR}_{\frac{1}{2}^+}}},\\
& & \frac{\eta^b_0}{\bar{\Lambda}_{(s)}} = \frac{{\cal SR}_{\eta^b_0}}{\sqrt{{\cal SR}_{\frac{1}{2}^-} \times {\cal SR}_{\frac{3}{2}^+}}},\\
& & \frac{\eta^c_0}{\bar{\Lambda}^{3/2}_{(s)}} = \frac{{\cal SR}_{\eta^c_0}}{\sqrt{{\cal SR}_{\frac{1}{2}^-} \times {\cal SR}_{\frac{3}{2}^+}}}.\label{eta0c}
\end{eqnarray}

As mentioned in Ref.\cite{HQEFTSR1,HQEFTSR2}, the QCD higher order corrections are not included in the above formulae. As far as the determination of universal wave functions is concerned, the effects of radiative corrections are expected to be largely cancelled in the ratios of sum rules for wave functions and decay constants (c.f. Eqs.(\ref{zeta})-(\ref{eta0c})) and therefore not influence the final results significantly.

For the $B_{(s)} \rightarrow D^{1/2}_{(s)}$ decays, the contributions from $\chi^{b(c)}_{1(2)}$ are generally expected to be very small and can be safely neglected, supported by the relativistic quark model and QCD sum rule study\cite{HQEFTSR2,cmos1,cmos2,cmos3,cmos4}.
However, as pointed out in Ref.\cite{HQEFTSR1}, under the condition $\eta^{b(c)}_i=0(i=1,2,3) $, the resulting branching fraction for the $B \rightarrow D^*_2 l \bar{\nu}_l$ decay seems to exceed the CLEO upper limit when including $1/m_Q$ contributions. Considering this, the wave functions $\eta^{b(c)}_i(i=1,2,3)$ may give significant contributions and require consideration for the $B_{(s)} \rightarrow D^{3/2}_{(s)}$ modes. The form factors depend on $\eta^b_i(i=1,2,3)$ only through their linear combination $\eta_b$ (c.f. Eq.(\ref{etab})). Adopting the assumption made in Ref.\cite{HQEFTSR1} that $\eta^{b(c)}_i(i=1,2,3)$ have a similar dependence on $q^2$ as $\tilde{\tau}$, we have
\begin{eqnarray}\label{eta}
\eta(q^2)= \frac{\eta(q^2_{max})}{1+\frac{q^2_{max}-q^2}{2 m_{B_{(s)}} m_{D^{**}_{(s)}} a^2}},
\end{eqnarray}
where $\eta = \eta^b, \eta^c_i (i =1,2,3)$, $q^2_{max} = ( m_{B_{(s)}} - m_{D^{**}_{(s)}})^2$, and $a^2=0.67$.

\section{Numerical results and discussions of the form factors}

With Eqs.(\ref{gP})-(\ref{eta}), we are now in a position to calculate the $B_{(s)} \rightarrow D^{**}_{(s)}$ form factors.
For the masses of heavy mesons, which have been well established in experiments, we use the latest values given by the particle data group (PDG)\cite{PDG2022},
\begin{eqnarray}
& & m_B=5.279\mbox{GeV}, \hspace{0.5cm} m_{B^*} = 5.325\mbox{GeV}, \hspace{0.5cm} m_{B_1} =5.726\mbox{GeV}, \hspace{0.5cm} m_{B^*_2}= 5.737\mbox{GeV},  \nonumber \\
& & m_{B_s}=5.367\mbox{GeV}, \hspace{0.5cm} m_{B^*_s} = 5.415\mbox{GeV}, \hspace{0.5cm} m_{B_{s1}}= 5.829\mbox{GeV}, \hspace{0.5cm} m_{B^*_{s2}}= 5.840\mbox{GeV}, \nonumber \\
& & m_D= 1.865\mbox{GeV}, \hspace{0.5cm} m_{D^*}= 2.007\mbox{GeV}, \hspace{0.5cm} m_{D^*_0} = 2.343\mbox{GeV}, \hspace{0.5cm} m_{D^\prime_1} = 2.412\mbox{GeV}, \nonumber \\
& & m_{D_1} = 2.422\mbox{GeV}, \hspace{0.5cm} m_{D^*_2} = 2.461\mbox{GeV}, \hspace{0.5cm} m_{D_s}= 1.968\mbox{GeV}, \hspace{0.5cm} m_{D^*_s} = 2.112\mbox{GeV}, \nonumber \\
& &  m_{D^*_{s0}}= 2.318\mbox{GeV}, \hspace{0.5cm} m_{D^\prime_{s1}} = 2.460\mbox{GeV}, \hspace{0.5cm} m_{D_{s1}}=2.535\mbox{GeV}, \hspace{0.5cm} m_{D^*_{s2}}= 2.569\mbox{GeV}. \nonumber \\
\end{eqnarray}
Furthermore, we adopt the masses estimated in the context of effective theory for the heavy mesons predicted by the quark model but not observed in experiments\cite{HQETCHPT},
\begin{eqnarray}
m_{B^*_0}= 5.681\mbox{GeV}, \hspace{0.5cm} m_{B^\prime_1}= 5.719\mbox{GeV}, \hspace{0.5cm} m_{B^*_{s0}}= 5.711\mbox{GeV}, \hspace{0.5cm}
m_{B^\prime_{s1}}= 5.756\mbox{GeV}.
\end{eqnarray}
For the masses of heavy quarks, we take the values\cite{HQEFTSR2}
\begin{eqnarray}
m_b = 4.67 \pm 0.05\mbox{GeV}, \hspace{0.5cm} m_c = 1.35 \pm 0.05\mbox{GeV}.
\end{eqnarray}
The condensates in Eqs.(\ref{SRzeta})-(\ref{SR3/2p}) have the typical values\cite{HQEFTSR2,conden}
\begin{eqnarray}
& & \langle \bar{q}q \rangle = \left\{
       \begin{array}{ll}
       -(0.23\mbox{GeV})^3, & \mbox{for $q=u, d$}  \\
      -0.8(0.23\mbox{GeV})^3, & \mbox{for $q=s$}
           \end{array}
         \right., \\
& & i \langle \bar{q} \sigma_{\alpha \beta} T^a F^{a \alpha \beta} q \rangle = - m^2_0 \langle \bar{q} q \rangle, \hspace{0.4cm} \mbox{with} \hspace{0.4cm} m^2_0 = 0.8 \mbox{GeV}^2, \\
& & \alpha_s \langle F^a_{\alpha \beta} F^{a \alpha \beta} \rangle =0.04 \mbox{GeV}^4.
\end{eqnarray}
For the values of $\eta^b, \eta^c_i(i=1,2,3)$ at $q^2= q^2_{max}$ in Eqs.(\ref{eta}), based on the analysis in Ref.\cite{HQEFTSR1}, we choose
\begin{eqnarray}
& & \eta^b(q^2_{max})= -0.6 \pm 0.1 {\rm GeV}^2, \hspace{0.5cm} \eta^c_1 (q^2_{max}) = 0.0 \pm 0.1 {\rm GeV}^2, \nonumber \\
& & \eta^c_2(q^2_{max})= -0.6 \pm 0.1 {\rm GeV}^2,  \hspace{0.5cm} \eta^c_3(q^2_{max})=0.4 \pm 0.1 {\rm GeV}^2,
\end{eqnarray}
 for which the branching fractions for the $B_{(s)} \rightarrow D^*_{(s)2} l \bar{\nu}_l$ decays can be significantly suppressed and the corresponding results for decays with $D_{(s)1}$ in the final states are largely unaffected.

\begin{figure}
\centering
\includegraphics[width=3in]{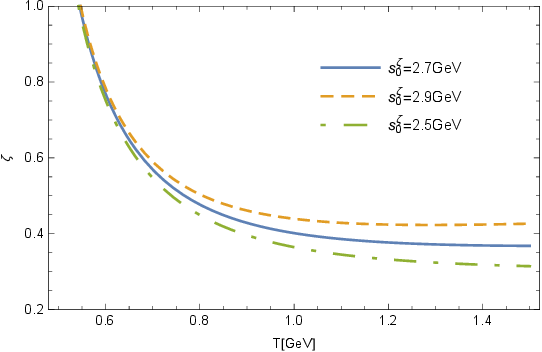} \hfill
\includegraphics[width=3in]{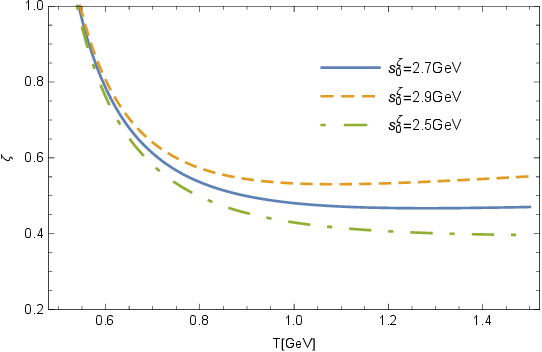} \\
\vspace{0.5cm}
\includegraphics[width=3in]{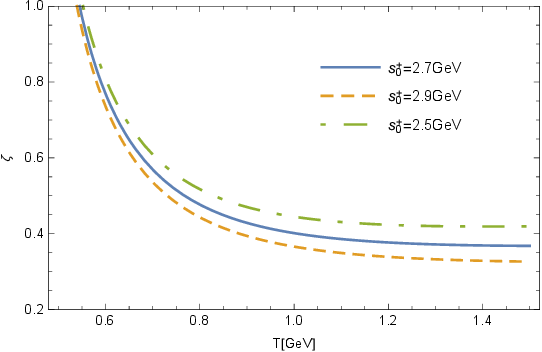}\hfill
\includegraphics[width=3in]{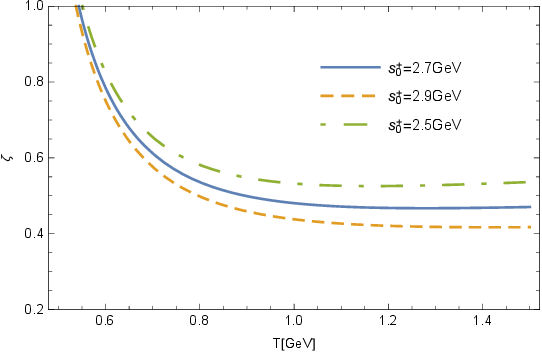} \\
\vspace{0.5cm}
\includegraphics[width=3in]{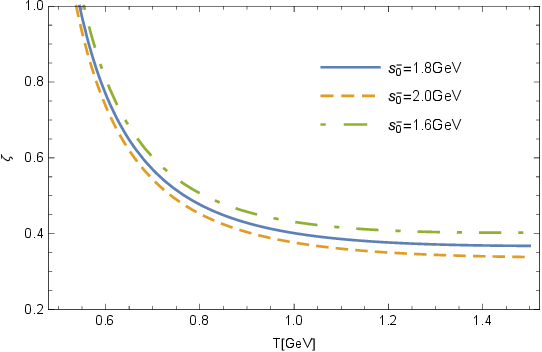}\hfill
\includegraphics[width=3in]{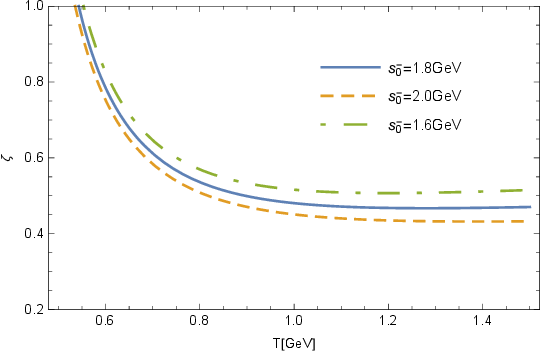}
\caption{Variation in $\zeta$ with respect to $T$ for different `$s_0$' parameters at $q^2=0$ (left column) and $q^2_{max}$ (right column).} \label{fig1}
\end{figure}

\begin{figure}
\centering
\includegraphics[width=3in]{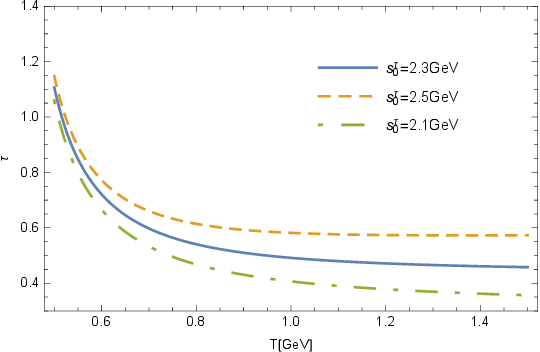} \hfill
\includegraphics[width=3in]{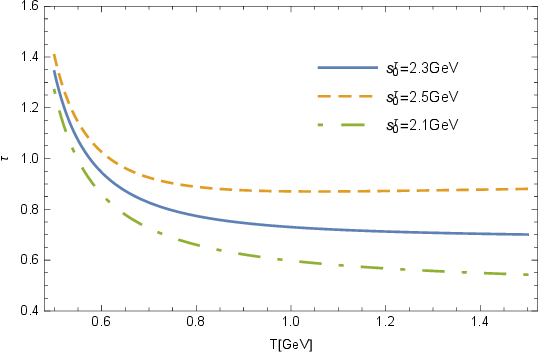} \\
\vspace{0.5cm}
\includegraphics[width=3in]{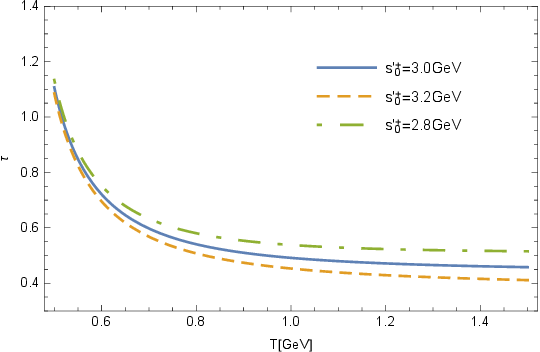}\hfill
\includegraphics[width=3in]{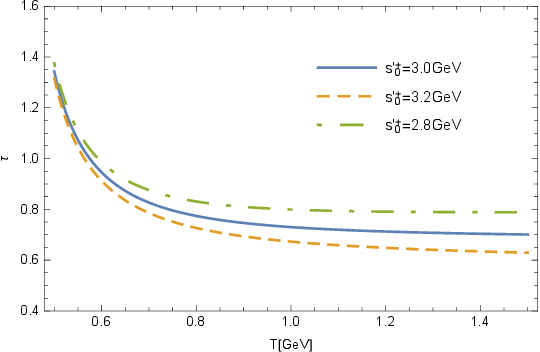} \\
\vspace{0.5cm}
\includegraphics[width=3in]{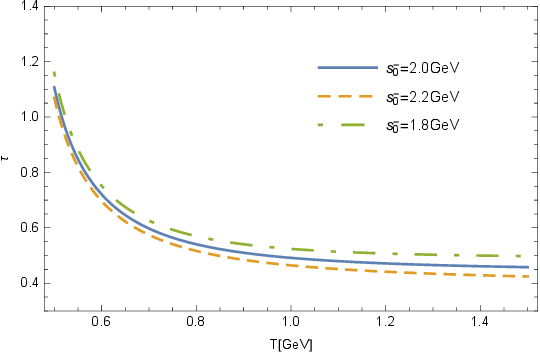}\hfill
\includegraphics[width=3in]{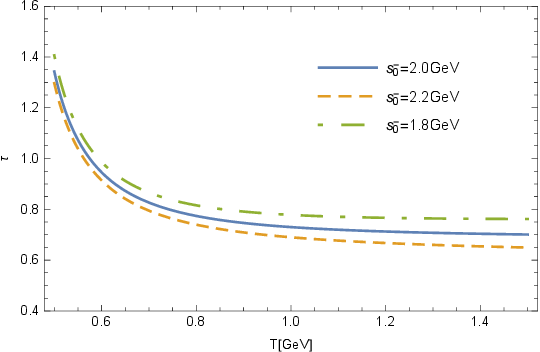}
\caption{Variation in $\tau$ with respect to $T$ for different `$s_0$' parameters at $q^2=0$ (left column) and $q^2_{max}$ (right column).} \label{fig2}
\end{figure}

\begin{figure}
\centering
\includegraphics[width=3in]{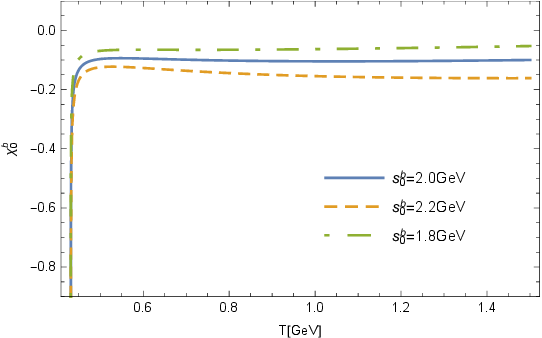} \hfill
\includegraphics[width=3in]{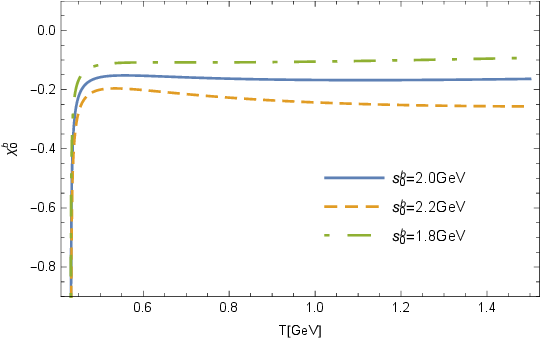} \\
\vspace{0.5cm}
\includegraphics[width=3in]{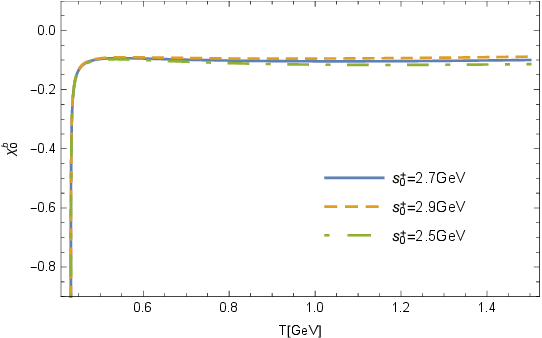}\hfill
\includegraphics[width=3in]{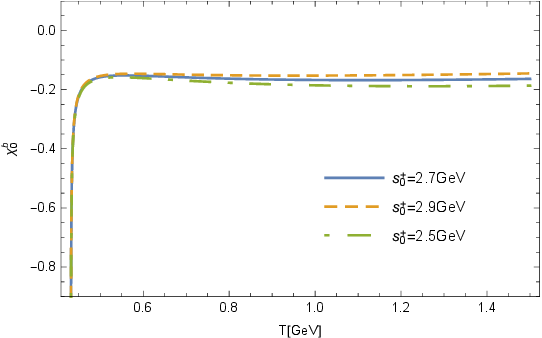} \\
\vspace{0.5cm}
\includegraphics[width=3in]{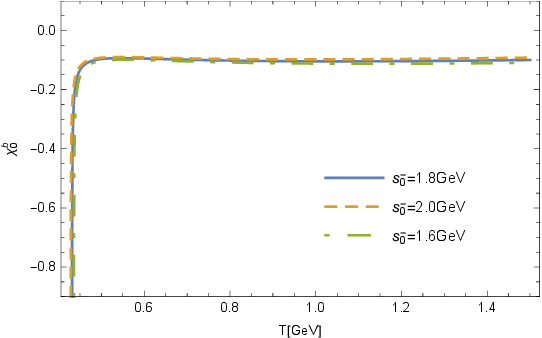}\hfill
\includegraphics[width=3in]{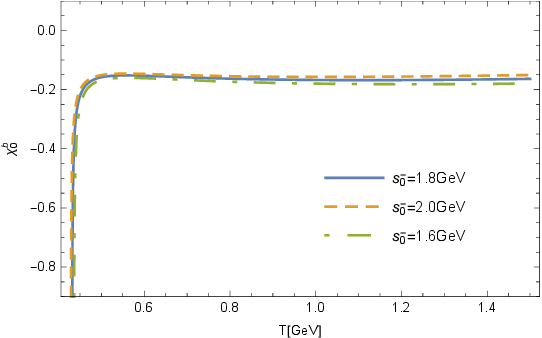}
\caption{Variation in $\chi^b_0$ with respect to $T$ for different `$s_0$' parameters at $q^2=0$ (left column) and $q^2_{max}$ (right column).} \label{fig3}
\end{figure}

\begin{figure}
\centering
\includegraphics[width=3in]{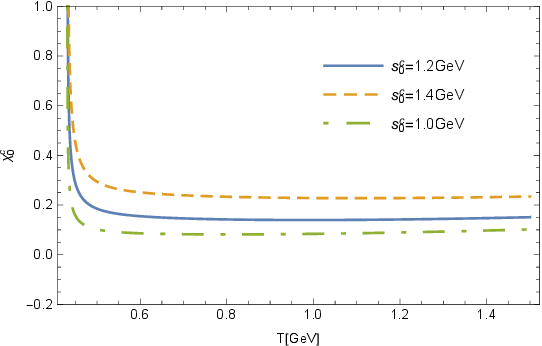} \hfill
\includegraphics[width=3in]{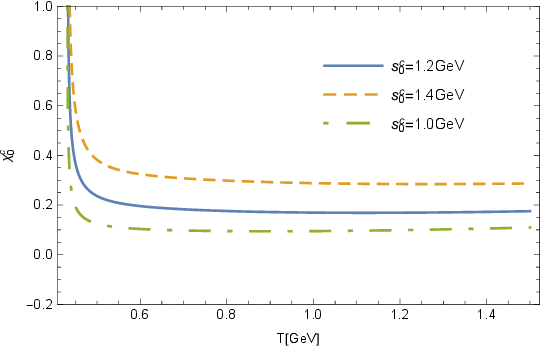} \\
\vspace{0.5cm}
\includegraphics[width=3in]{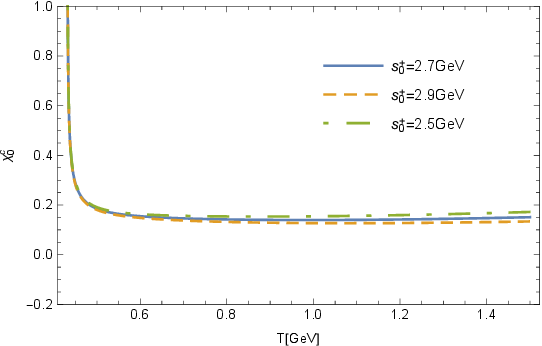}\hfill
\includegraphics[width=3in]{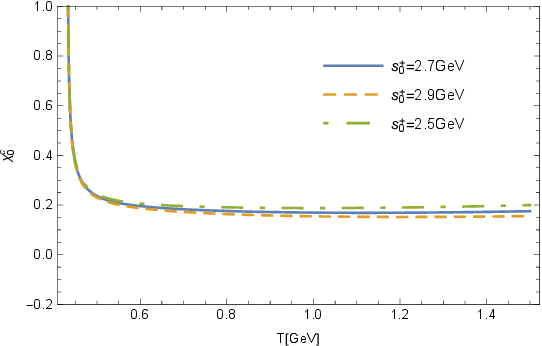} \\
\vspace{0.5cm}
\includegraphics[width=3in]{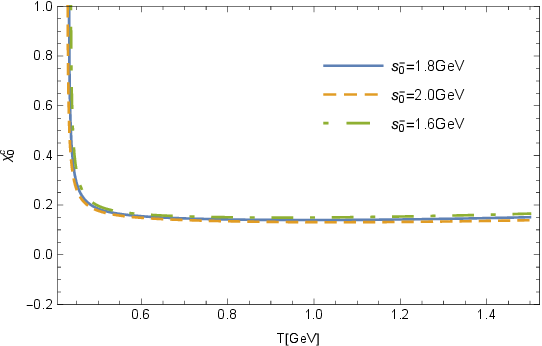}\hfill
\includegraphics[width=3in]{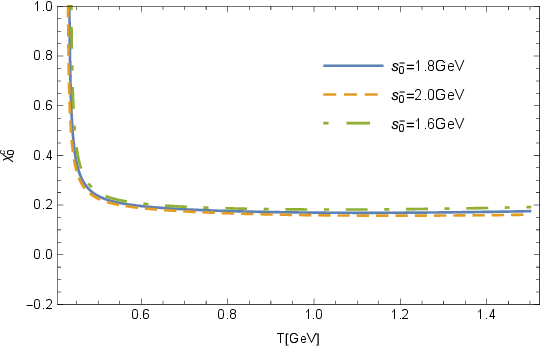}
\caption{Variation in $\chi^c_0$ with respect to $T$ for different `$s_0$' parameters at $q^2=0$ (left column) and $q^2_{max}$ (right column). } \label{fig4}
\end{figure}

\begin{figure}
\centering
\includegraphics[width=3in]{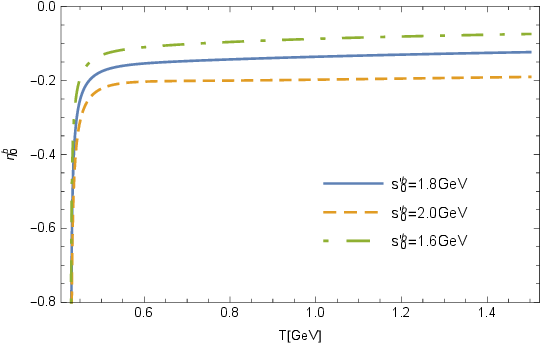} \hfill
\includegraphics[width=3in]{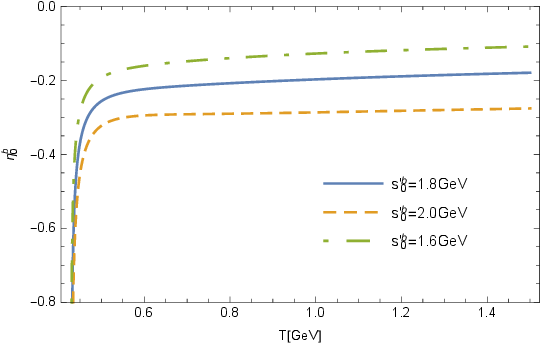} \\
\vspace{0.5cm}
\includegraphics[width=3in]{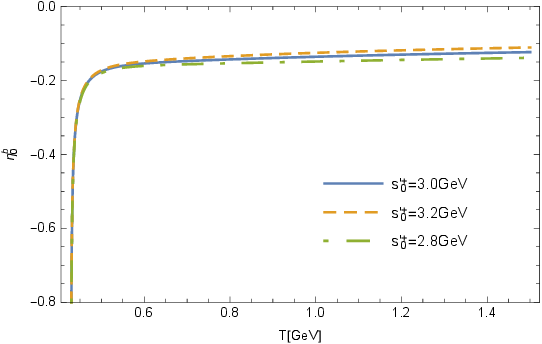}\hfill
\includegraphics[width=3in]{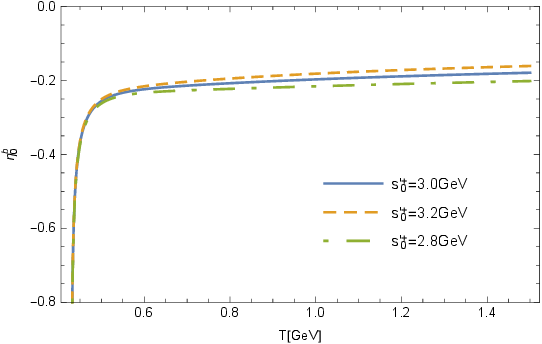} \\
\vspace{0.5cm}
\includegraphics[width=3in]{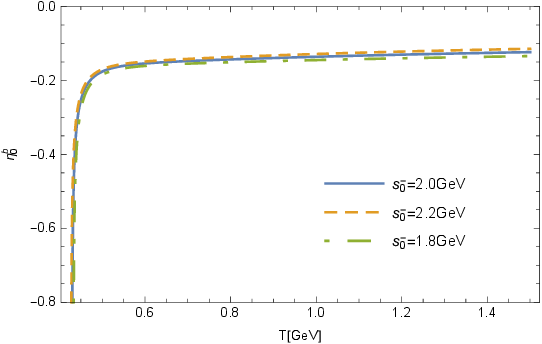}\hfill
\includegraphics[width=3in]{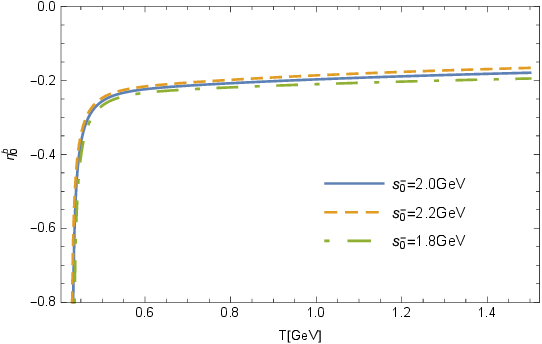}
\caption{Variation in $\eta^b_0$ with respect to $T$ for different `$s_0$' parameters at $q^2=0$ (left column) and $q^2_{max}$ (right column). } \label{fig5}
\end{figure}

\begin{figure}
\centering
\includegraphics[width=3in]{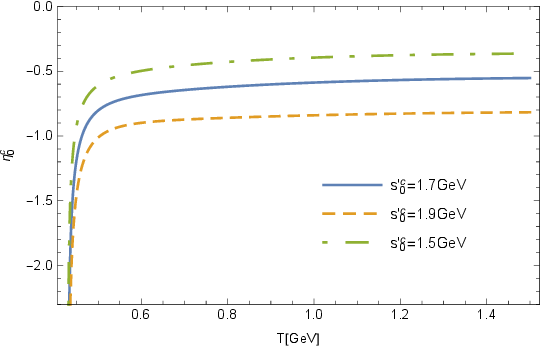} \hfill
\includegraphics[width=3in]{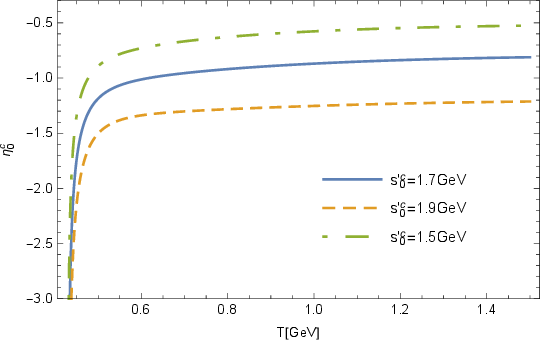} \\
\vspace{0.5cm}
\includegraphics[width=3in]{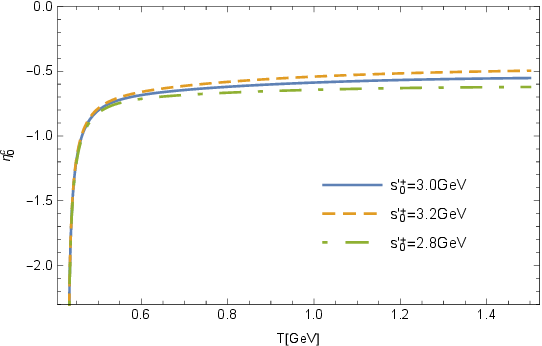}\hfill
\includegraphics[width=3in]{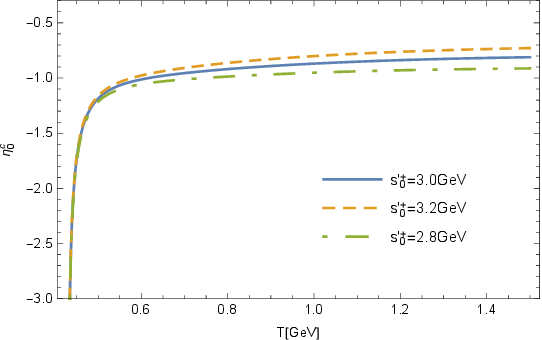} \\
\vspace{0.5cm}
\includegraphics[width=3in]{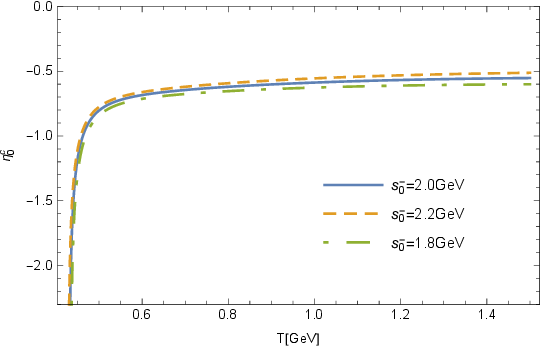}\hfill
\includegraphics[width=3in]{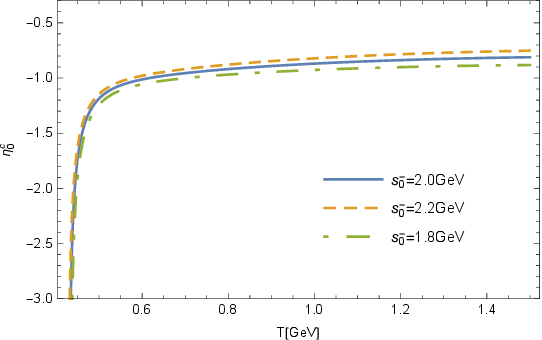}
\caption{Variation in $\eta^c_0$ with respect to $T$ for different `$s_0$' parameters at $q^2=0$ (left column) and $q^2_{max}$ (right column). } \label{fig6}
\end{figure}

From Eqs.(\ref{SRzeta})-(\ref{eta0c}), it is easily observed that the sum rules for the universal wave functions $\zeta$, $\tau$, $\chi^{b(c)}_0$, $\eta^{b(c)}_0$ contain ten free parameters $s^\zeta_0$, $s^\tau_0$, $s^b_0$, $s^c_0$, $s^{\prime b}_0$, $s^{\prime c}_0$, $s^+_0$, $s^{\prime +}_0$, $s^-_0$, and $T$, where the `$s_0$' parameters are related to the threshold energies of initial and final heavy mesons, and $T$ is the Borel parameter.
The allowed regions for these free parameters are determined by requiring the curves of the wave functions $\zeta$, $\tau$, $\chi^{b(c)}_0$, and $\eta^{b(c)}_0$ to be most stable. In practice, we adjust the free parameters for all relevant $B_{(s)} \rightarrow D^{1/2+}_{(s)}$ form factors consistently, and the same procedure is performed for those of the $B_{(s)} \rightarrow D^{3/2+}_{(s)}$ decays. The variations in the wave functions $\zeta$, $\tau$, $\chi^{b(c)}_0$, and $\eta^{b(c)}_0$ with respect to $T$ for different `$s_0$' parameters at $q^2=0, q^2_{max}$ are illustrated in FIG.\ref{fig1}-\ref{fig6}. For each plot, the relevant `$s_0$' parameters not displayed in the legend are fixed at their central values shown in other rows of the same figure. Based on these curves, it is reasonable to choose
\begin{eqnarray}
& & s^\zeta_0 = 2.7 \pm 0.2{\rm GeV}, \hspace{0.5cm} s^b_0 = 2.0 \pm 0.2{\rm GeV}, \hspace{0.5cm} s^c_0 = 1.2 \pm 0.2 {\rm GeV}, \nonumber \\
& & s^+_0 = 2.7 \pm 0.2{\rm GeV}, \hspace{0.5cm} s^-_0 = 1.8 \pm 0.2{\rm GeV}, \hspace{0.5cm} T = 1.2 \pm 0.2{\rm GeV},
\end{eqnarray}
for the $B_{(s)} \rightarrow D^{1/2}_{(s)}$ decays and
\begin{eqnarray}
& & s^\tau_0 = 2.3 \pm 0.2{\rm GeV}, \hspace{0.5cm} s^{\prime b}_0 = 1.8 \pm 0.2{\rm GeV}, \hspace{0.5cm} s^{\prime c}_0 = 1.7 \pm 0.2 {\rm GeV}, \nonumber \\
& & s^{\prime +}_0 = 3.0 \pm 0.2{\rm GeV}, \hspace{0.5cm} s^-_0 = 2.0 \pm 0.2{\rm GeV}, \hspace{0.5cm} T = 1.1 \pm 0.2{\rm GeV},
\end{eqnarray}
for the $B_{(s)} \rightarrow D^{3/2}_{(s)}$ decays.

\begin{figure}
\centering
\includegraphics[width=3in]{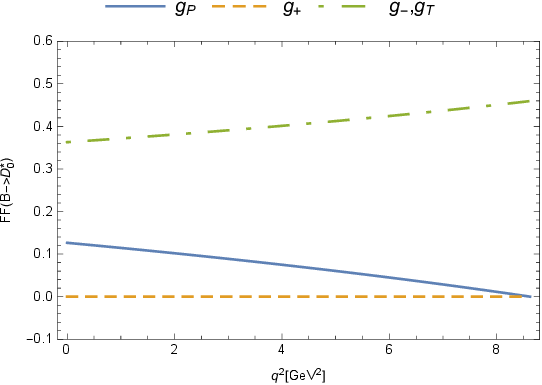} \hfill
\includegraphics[width=3in]{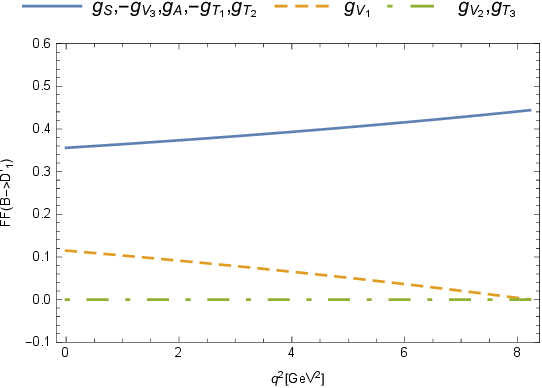} \\
\vspace{0.5cm}
\includegraphics[width=3in]{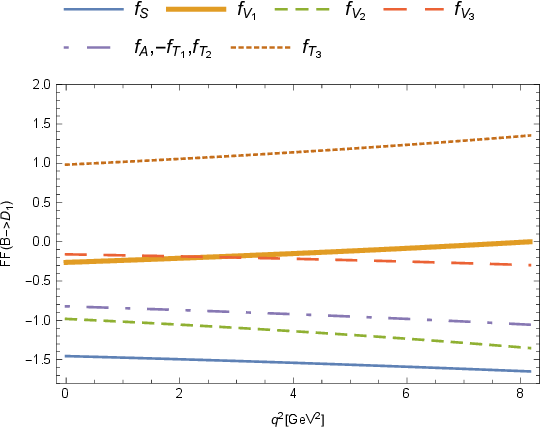} \hfill
\includegraphics[width=3in]{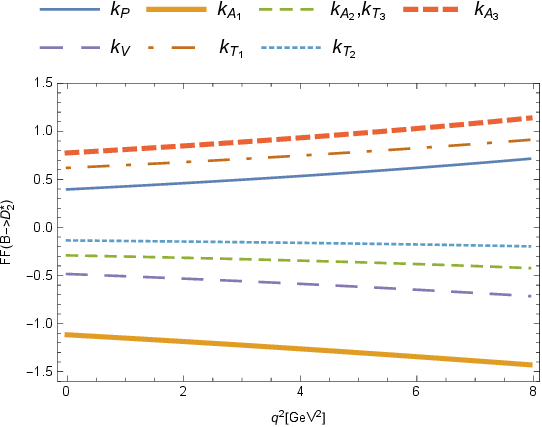} \\
\caption{Form factors of $B \rightarrow D^{**}$ decays as functions of $q^2$.} \label{fig7}
\end{figure}

With the above considerations, we calculate all the relevant form factors of the $B_{(s)} \rightarrow D^{**}_{(s)}$ decays systematically. For the case of the $B \rightarrow D^{**}$ modes, the form factors as functions of $q^2$ in the entire physical region are shown in FIG.\ref{fig7}.
The form factors of the $B_s \rightarrow D^{**}_s$ decays as functions of $q^2$ have similar behaviors. The numerical results of the form factors at $q^2=0, q^2_{max}$ are presented in TABLE~\ref{table1}-\ref{table4}. For the form factors of the $B_{(s)} \rightarrow D^{1/2+}_{(s)}$ decays, the uncertainties originate from the `$s_0$' parameters, Borel parameter $T$, and heavy quark masses $m_Q(Q=b,c)$, which are approximately $20\%-30\%$ added in quadrature. For the form factors of the $B_{(s)} \rightarrow D^{3/2+}_{(s)}$ decays, additional uncertainties induced by the values of $\eta^b, \eta^c_i(i=1,2,3)$ at $q^2= q^2_{max}$ are also included, and the maximum total uncertainties can reach $90\%$. Note that the uncertainties quoted here merely indicate variations in our results within the chosen ranges of the relevant parameters mentioned above. The form factors $g_P$, $g_{V_1}$, and $f_{V_1}$ approach zero at $q^2= q^2_{max}$. The form factor $g_+$ is equal to zero in the entire $q^2$ region up to the next leading order of heavy quark expansion. In addition, the form factors $g_{V_2}=-g_{T_3}=0$ and $g_S=-g_{V_3} = g_A$ under the condition that the contributions from chromomagnetic operators are neglected. The uncertainties of $k_{A_2}$, $k_{T_2}$, and $k_{T_3}$ are considerably smaller because these form factors only depend on the wave functions $\eta^b$ and $\eta^c_2$ given by Eq.(\ref{eta}). Owing to the approximate $SU(3)$ flavor symmetry, the form factors of $B \rightarrow D^{**}$ decays are very close to their strange counterparts. A comparison of the $B \rightarrow D^{**}$ form factors at $q^2=q^2_{max}$ from this study with those of other groups is shown in TABLE~\ref{compartable1}-\ref{compartable4}, where the original values from LFQM\cite{LFQM1,LFQM2}, ISGW2\cite{ISGW21}, and LCSRB\cite{LCSRB1} are converted to meet the current definitions of form factors using the formulae in APPENDIX~\ref{ConFor}. We can see that large differences exist among the form factors given by different groups. Overall, our results are in better agreement with the values obtained in HQET with available experimental measurements as inputs, i.e. the HQET+EXP. method\cite{HQETE3,HQETE4}.
\begin{table}
\centering
\begin{tabular}{|c|c|c|c|c|}
\hline Decays & $q^2$ & $g_P$ & $g_+$ & $g_-=g_T$ \\
\hline $B \rightarrow D^*_0$ & $0$ & $0.13^{+0.03}_{-0.03}$ & $0$ & $0.36^{+0.08}_{-0.07}$ \\
\cline{2-5} & $q^2_{max}$ & $0 $ & $0$  & $0.46^{+0.10}_{-0.10}$ \\
\hline
$B_s \rightarrow D^*_{s0}$ & $0$ & $0.14^{+0.03}_{-0.03}$ &  $0$ & $0.36^{+0.08}_{-0.07}$ \\
\cline{2-5} & $q^2_{max}$ & $0 $ & $0$  & $0.47^{+0.11}_{-0.10}$ \\
\hline
\end{tabular}
\caption{Form factors of $B_{(s)} \rightarrow D^*_{(s)0}$ decays at $q^2=0, q^2_{max}$.} \label{table1}
\end{table}
\begin{table}
\centering
\begin{tabular}{|c|c|c|c|c|}
\hline Decays & $q^2$ & $g_S=-g_{V_3}=g_A=-g_{T_1}=g_{T_2}$ & $g_{V_1}$ & $g_{V_2}= -g_{T_3}$ \\
\hline $B \rightarrow D^\prime_1$ & $0$ & $0.36^{+0.08}_{-0.07}$ & $0.11^{+0.03}_{-0.02}$ & $0$ \\
\cline{2-5} & $q^2_{max}$ & $0.44^{+0.10}_{-0.09} $ & $0$  & $0$ \\
\hline
$B_s \rightarrow D^\prime_{s1}$ & $0$ & $0.35^{+0.08}_{-0.07}$ &  $0.11^{+0.03}_{-0.02}$ & $0$ \\
\cline{2-5} & $q^2_{max}$ & $0.44^{+0.10}_{-0.10} $ & $0$  & $0$ \\
\hline
\end{tabular}
\caption{Form factors of $B_{(s)} \rightarrow D^\prime_{(s)1}$ decays at $q^2=0, q^2_{max}$.} \label{table2}
\end{table}
\begin{table}
\centering
\begin{tabular}{|c|c|c|c|c|c|c|c|}
\hline Decays & $q^2$ & $f_S$  & $f_{V_1}$ & $f_{V_2}$ & $f_{V_3}$  & $f_A=-f_{T_1} = f_{T_2}$ & $f_{T_3}$ \\
\hline $B \rightarrow D_1$ & $0$ & $-1.45^{+0.38}_{-0.43}$ & $-0.26^{+0.05}_{-0.06}$ & $-0.98^{+0.21}_{-0.23}$ & $-0.16^{+0.15}_{-0.15}$ & $-0.82^{+0.15}_{-0.18}$ & $0.98^{+0.23}_{-0.21}$ \\
\cline{2-8} & $q^2_{max}$ & $-1.65^{+0.49}_{-0.54} $ & $0$  & $-1.35^{+0.31}_{-0.35}$ & $-0.30^{+0.22}_{-0.22}$ & $-1.06^{+0.20}_{-0.22}$ & $1.35^{+0.35}_{-0.31}$ \\
\hline
$B_s \rightarrow D_{s1}$ & $0$ & $-1.48^{+0.37}_{-0.42}$ & $-0.25^{+0.05}_{-0.05}$ & $-1.01^{+0.21}_{-0.24}$ & $-0.17^{+0.14}_{-0.15}$ & $-0.83^{+0.16}_{-0.19}$ & $1.01^{+0.24}_{-0.21}$ \\
\cline{2-8} & $q^2_{max}$ & $-1.69^{+0.47}_{-0.53} $ & $0$  & $-1.37^{+0.30}_{-0.36}$ & $-0.32^{+0.21}_{-0.21}$ & $-1.06^{+0.21}_{-0.23}$ & $1.37^{+0.36}_{-0.30}$ \\
\hline
\end{tabular}
\caption{Form factors of $B_{(s)} \rightarrow D_{(s)1}$ decays at $q^2=0, q^2_{max}$.} \label{table3}
\end{table}
\begin{table}
\centering
\begin{tabular}{|c|c|c|c|c|c|c|c|c|}
\hline Decays & $q^2$ & $k_P$ & $k_V$ & $k_{A_1}$ & $k_{A_2}=k_{T_3}$ & $k_{A_3}$ & $k_{T_1}$ & $k_{T_2}$ \\
\hline $B \rightarrow D^*_2$ & $0$ & $0.40^{+0.17}_{-0.16}$ & $-0.48^{+0.15}_{-0.18}$ & $-1.12^{+0.35}_{-0.41}$ & $-0.29^{+0.05}_{-0.05}$ & $0.78^{+0.18}_{-0.16}$ & $0.62^{+0.18}_{-0.15}$ & $-0.14^{+0.03}_{-0.02}$ \\
\cline{2-9} & $q^2_{max}$ & $0.71^{+0.27}_{-0.23} $ & $-0.71^{+0.23}_{-0.27}$ & $-1.43^{+0.44}_{-0.52}$ & $-0.42^{+0.07}_{-0.07}$ & $1.14^{+0.27}_{-0.24}$ & $0.91^{+0.26}_{-0.22}$ & $-0.20^{+0.04}_{-0.03}$ \\
\hline
$B_s \rightarrow D^*_{s2}$ & $0$ & $0.45^{+0.19}_{-0.16}$ & $-0.53^{+0.16}_{-0.19}$ &  $-1.21^{+0.36}_{-0.43}$ & $-0.27^{+0.04}_{-0.05}$ & $0.80^{+0.20}_{-0.16}$ & $0.65^{+0.19}_{-0.16}$ & $-0.12^{+0.02}_{-0.02}$ \\
\cline{2-9} & $q^2_{max}$ & $0.77^{+0.27}_{-0.23}$ & $-0.77^{+0.23}_{-0.27}$ & $-1.54^{+0.46}_{-0.53}$ & $-0.39^{+0.07}_{-0.06}$ & $1.15^{+0.28}_{-0.23}$ & $0.94^{+0.27}_{-0.22}$ & $-0.17^{+0.02}_{-0.03}$ \\
\hline
\end{tabular}
\caption{Form factors of $B_{(s)} \rightarrow D^*_{(s)2}$ decays at $q^2=0, q^2_{max}$.} \label{table4}
\end{table}

\begin{table}
\centering
\begin{tabular}{|c|c|c|c|c|}
\hline   Ref. & $g_P$ & $g_+$ & $g_-$ & $g_T$ \\
\hline  This study & $0 $ & $0$  & $0.46^{+0.10}_{-0.10}$ & $0.46^{+0.10}_{-0.10}$\\
\hline   HQET+EXP.\cite{HQETE3,HQETE4}& $0.20^{+0.06}_{-0.06}$ & $-0.18^{+0.05}_{-0.05}$ & $0.70^{+0.21}_{-0.21}$ & $0.80^{+0.24}_{-0.24}$ \\
\hline    LFQM\cite{LFQM1} & ---& $-0.22$ & $0.25$ & ---\\
\hline    ISGW2\cite{ISGW21} &--- & $0.01$ & $0.60$ &--- \\
\hline
\end{tabular}
\caption{Comparison of the $B \rightarrow D^*_0$ form factors at $q^2=q^2_{max}$ from this study with those of other groups.} \label{compartable1}
\end{table}
\begin{table}
\centering
\resizebox{\linewidth}{!}{
\begin{tabular}{|c|c|c|c|c|c|c|c|c|}
\hline   Ref. & $g_S$ & $g_{V_1}$ & $g_{V_2}$ &  $g_{V_3}$ & $g_A$ & $g_{T_1}$ & $g_{T_2}$ & $g_{T_3}$  \\
\hline   This study & $0.44^{+0.10}_{-0.09} $ & $0$  & $0$ & $-0.44^{+0.09}_{-0.10} $ & $0.44^{+0.10}_{-0.09} $ & $-0.44^{+0.09}_{-0.10} $ & $0.44^{+0.10}_{-0.09} $ & $0$\\
\hline   HQET+EXP.\cite{HQETE3,HQETE4} & $0.61^{+0.19}_{-0.18}$ & $0.01^{+0.01}_{-0.00}$ & $0.31^{+0.18}_{-0.17}$ & $-1.02^{+0.35}_{-0.34}$ & $0.71^{+0.21}_{-0.22}$ & $-0.61^{+0.08}_{-0.09}$ & $0.79^{+0.23}_{-0.24}$ & $0.31^{+0.18}_{-0.17}$ \\
\hline   LFQM\cite{LFQM1} & ---& $-0.28$ & $1.31$ & $-0.47$ & $-0.13$ & ---&--- & ---\\
\hline   ISGW2\cite{ISGW21} & ---& $0.01$ & $0.60$ & $-0.05$  & $-0.19$ &--- &--- &--- \\
\hline
\end{tabular}
}
\caption{Comparison of the $B \rightarrow D^\prime_1$ form factors at $q^2=q^2_{max}$ from this study with those of other groups.}\label{compartable2}
\end{table}
\begin{table}
\centering
\resizebox{\linewidth}{!}{
\begin{tabular}{|c|c|c|c|c|c|c|c|c|}
\hline   Ref. & $f_S$  & $f_{V_1}$ & $f_{V_2}$ & $f_{V_3}$  & $f_A$ & $f_{T_1}$ & $f_{T_2}$ & $f_{T_3}$ \\
\hline  This study & $-1.65^{+0.49}_{-0.54} $ & $0$  & $-1.35^{+0.31}_{-0.35}$ & $-0.30^{+0.22}_{-0.22}$ & $-1.06^{+0.20}_{-0.22}$ & $1.06^{+0.22}_{-0.20}$ & $-1.06^{+0.20}_{-0.22}$  & $1.35^{+0.35}_{-0.31}$ \\
\hline  HQET+EXP.\cite{HQETE3,HQETE4} & $-1.31^{+0.13}_{-0.13}$ & $-0.34^{+0.04}_{-0.03}$ & $-2.21^{+0.65}_{-0.66}$ & $1.24^{+0.62}_{-0.62}$ & $-0.74^{+0.07}_{-0.07}$ & $0.40^{+0.04}_{-0.04}$ & $-0.74^{+0.07}_{-0.07}$ & $-0.54^{+0.89}_{-0.89}$ \\
\hline  LFQM\cite{LFQM1} &--- & $1.10$ & $-0.80$ & $0.53$ & $0.31$ &--- &--- &--- \\
\hline  ISGW2\cite{ISGW21}  &--- & $0.69$ & $-0.32$ & $0.32$ & $0.18$ &--- & ---& ---\\
\hline
\end{tabular}
}
\caption{Comparison of the $B \rightarrow D_1$ form factors at $q^2=q^2_{max}$ from this study with those of other groups.}\label{compartable3}
\end{table}
\begin{table}
\centering
\resizebox{\linewidth}{!}{
\begin{tabular}{|c|c|c|c|c|c|c|c|c|}
\hline   Ref. & $k_P$ & $k_V$ & $k_{A_1}$ & $k_{A_2}$ & $k_{A_3}$ & $k_{T_1}$ & $k_{T_2}$ & $k_{T_3}$\\
\hline   This study & $0.71^{+0.27}_{-0.23} $ & $-0.71^{+0.23}_{-0.27}$ & $-1.43^{+0.44}_{-0.52}$ & $-0.42^{+0.07}_{-0.07}$ & $1.14^{+0.27}_{-0.24}$ & $0.91^{+0.26}_{-0.22}$ & $-0.20^{+0.04}_{-0.03}$ & $-0.42^{+0.07}_{-0.07}$\\
\hline  HQET+EXP.\cite{HQETE3,HQETE4} & $1.05^{+0.32}_{-0.33}$ & $0.20^{+0.46}_{-0.47}$ & $-1.40^{+0.14}_{-0.14}$ & $0.26^{+0.15}_{-0.16}$ & $0.06^{+0.49}_{-0.49}$ & $0.70^{+0.07}_{-0.07}$ & $0.87^{+0.18}_{-0.17}$ & $-0.26^{+0.16}_{-0.15}$ \\
\hline  LFQM\cite{LFQM1} &--- & $-0.06$ & $-0.85$ & $2.12$ & $-0.93$ & ---&--- & ---\\
\hline  ISGW2\cite{ISGW21} &--- & $-0.04$ & $-0.52$ & $1.27$ & $-0.56$ &--- &--- &--- \\
\hline  LCSRB\cite{LCSRB1} & ---& $2.04^{+0.36}_{-0.42}$ & $3.29^{+0.53}_{-0.61}$ & $-0.80^{+2.36}_{-2.44}$ & $-1.04^{+1.23}_{-1.31}$ & $1.67^{+0.31}_{-0.33}$ & $0.17^{+1.17}_{-1.17}$ & $-0.28^{+2.07}_{-2.10}$\\
\hline  LFQM\cite{LFQM2} & ---& $1.06^{+0.19}_{-0.22}$ & $1.80^{+0.33}_{-0.34}$ & $-0.15^{+1.48}_{-1.42}$ & $-0.72^{+0.72}_{-0.75}$ & $0.89^{+0.15}_{-0.17}$ & $-0.01^{+0.62}_{-0.64}$ & $0.01^{+1.14}_{-1.09}$\\
\hline
\end{tabular}
}
\caption{Comparison of the $B \rightarrow D^*_2$ form factors at $q^2=q^2_{max}$ from this study with those of other groups.}\label{compartable4}
\end{table}

\section{$B_{(s)} \rightarrow D^{**}_{(s)} l \bar{\nu}_l$ decays and possible NP effects}

In this section, we apply the form factors obtained above to investigate the relevant $B_{(s)} \rightarrow D^{**}_{(s)} l \bar{\nu}_l$ decays and possible NP effects.

The effective Hamiltonian describing the $b \rightarrow c l \bar{\nu}_l$ transition with general NP contributions can be written as\cite{WilsonC}
\begin{eqnarray}\label{effH}
{\cal H}_{eff} = 2 \sqrt{2} G_F V_{cb} \left [ (1+ C_{V_L})O_{V_L} + C_{V_R}O_{V_R} + C_{S_L}O_{S_L} + C_{S_R}O_{S_R} + C_T O_T \right ],
\end{eqnarray}
with
\begin{eqnarray}
& & O_{S_L} = ( \bar{c}  P_L b ) ( \bar{l}  P_L \nu_l ), \hspace{1.3cm} O_{S_R} = ( \bar{c}  P_R b ) ( \bar{l}  P_L \nu_l ), \\
& & O_{V_L} = ( \bar{c} \gamma^\mu P_L b ) ( \bar{l} \gamma_\mu P_L \nu_l ), \hspace{0.5cm} O_{V_R} = ( \bar{c} \gamma^\mu P_R b ) ( \bar{l} \gamma_\mu P_L \nu_l ),\\
& & O_T = ( \bar{c} \sigma^{\mu \nu} P_L b ) ( \bar{l} \sigma_{\mu \nu} P_L \nu_l ),
\end{eqnarray}
where $P_{L(R)} = (1-(+)\gamma^5)/2$, and the active neutrinos are assumed to be left-handed. The SM corresponds to the Wilson coefficients $C_{S_{L(R)}}= C_{V_{L(R)}}=C_T=0$. The operators $O_{S_{L(R)}}$, $O_{V_{L(R)}}$, and $O_T$
represent the contributions from possible (pseudo-)scalar, (axial-)vector, and tensor interactions, respectively.

With this effective Hamiltonian, the differential decay widths w.r.t. $q^2$ for the $B_{(s)} \rightarrow D^{**}_{(s)} l \bar{\nu}_l$ decays can be obtained. Concretely\cite{HQETE4}, for the $B_{(s)} \rightarrow D^*_{(s)0} l \bar{\nu}_l$ decays,
\begin{eqnarray}
& & \frac{d \Gamma^{\rm SM}}{d q^2}= \frac{2 \Gamma_0 r^3 \sqrt{\omega^2-1}}{ m_{B_{(s)}} m_{D^*_{(s)0}}}  \frac{(\hat{q}^2 - \rho_l)^2}{\hat{q}^6} \left \{ g^2_- ( \omega -1) \left [ \rho_l [ (1+r^2)( 2 \omega -1) + 2 r ( \omega -2)] \right. \right. \nonumber \\
& & \hspace{1.5cm} \left. + (1-r)^2 ( \omega +1) \hat{q}^2 \right ] + g^2_+ ( \omega +1) \left [  \rho_l [ (1+r^2)( 2 \omega +1) - 2 r ( \omega +2)] \right. \nonumber \\
& & \hspace{1.5cm} \left. \left. +(1+r)^2 ( \omega -1) \hat{q}^2 \right ] - 2 g_- g_+ (1-r^2) ( \omega^2-1) ( \hat{q}^2 + 2 \rho_l ) \right \}, \\
& & \frac{d \Gamma}{d q^2} =  \frac{d \Gamma^{\rm SM}}{d q^2} ( 1+ C_{V_L} - C_{V_R} )^2 + \frac{ \Gamma_0 r^3 \sqrt{\omega^2-1}}{ m_{B_{(s)}} m_{D^*_{(s)0}}} \frac{(\hat{q}^2 - \rho_l)^2}{\hat{q}^4} \left \{ 3 (C_{S_R}-C_{S_L})^2 g^2_P \hat{q}^2 \right. \nonumber \\
& & \hspace{1.5cm} + 6 (C_{S_R}-C_{S_L}) g_P \sqrt{\rho_l} ( 1+ C_{V_L} - C_{V_R} ) \left [ g_- ( 1+r) ( \omega-1) \right. \nonumber \\
& & \hspace{1.5cm} \left. - g_+ ( 1-r) ( \omega +1) \right ] + 8 C_T g_T ( \omega^2 -1) \left [ 2 C_T g_T ( \hat{q}^2 + 2 \rho_l ) \right. \nonumber \\
& & \hspace{1.5cm} \left. \left. + 3 \sqrt{\rho_l} ( 1+ C_{V_L} - C_{V_R} ) [ g_+ (1+r) - g_- (1-r) ] \right ] \right \}.
\end{eqnarray}
For the $B_{(s)} \rightarrow D^\prime_{(s)1} l \bar{\nu}_l$ decays,
\begin{eqnarray}
& & \frac{d \Gamma^{\rm SM}}{d q^2}= \frac{ \Gamma_0 r^3 \sqrt{\omega^2-1}}{ m_{B_{(s)}} m_{D^\prime_{(s)1}}}  \frac{(\hat{q}^2 - \rho_l)^2}{\hat{q}^6} \left \{
g^2_{V_1} \left [ 2 \hat{q}^2 [ ( \omega -r)^2 + 2 \hat{q}^2 ] + \rho_l [ 4 ( \omega-r)^2 - \hat{q}^2 ] \right ] \right. \nonumber \\
& & \hspace{1.5cm} + ( \omega^2 -1) \left ( g^2_{V_2} \left [ 2 r^2 \hat{q}^2 ( \omega^2-1) + \rho_l [ 3 \hat{q}^2 + 4 r^2 ( \omega^2-1)]\right ]  \right. \nonumber \\
& & \hspace{1.5cm} + g^2_{V_3} \left [ 2 \hat{q}^2 ( \omega^2-1) + \rho_l [ 4 ( \omega-r)^2 - \hat{q}^2 ] \right ] + 2 g^2_A \hat{q}^2 ( 2 \hat{q}^2 + \rho_l) \nonumber \\
& & \hspace{1.5cm} + 2 g_{V_1} g_{V_2} \left [ 2 r \hat{q}^2 ( \omega-r) + \rho_l ( 3- r^2 -2r \omega) \right ] + 4 g_{V_1} g_{V_3} ( \omega -r) ( \hat{q}^2 + 2 \rho_l ) \nonumber \\
& & \hspace{1.5cm} \left. \left. + 2 g_{V_2} g_{V_3} \left [ 2r \hat{q}^2 ( \omega^2-1) + \rho_l [ 3 \omega \hat{q}^2 + 4 r ( \omega^2-1)] \right ] \right) \right\}, \label{DWD1pSM}\\
& & \frac{d \Gamma}{d q^2}=\frac{d \Gamma^{\rm SM}}{d q^2}( 1+ C_{V_L} - C_{V_R} )^2 + \frac{ \Gamma_0 r^3 \sqrt{\omega^2-1}}{ m_{B_{(s)}} m_{D^\prime_{(s)1}}} \frac{(\hat{q}^2 - \rho_l)^2}{\hat{q}^6} \left \{ 3 (C_{S_L} + C_{S_R} )^2  \right. \nonumber \\
& & \hspace{1.5cm} \times g^2_S ( \omega^2-1) \hat{q}^4 -6 (C_{S_L} + C_{S_R} ) ( 1+ C_{V_L} + C_{V_R} ) g_S ( \omega^2 -1) \hat{q}^2 \sqrt{\rho_l} \left [ g_{V_1} \right. \nonumber \\
& & \hspace{1.5cm} \left. + g_{V_2} ( 1- r \omega ) + g_{V_3} ( \omega -r ) \right ] + 16 C_T^2 ( \hat{q}^2 + 2 \rho_l) \left ( g^2_{T_1} \left [
\hat{q}^2 ( 2 + \omega^2) \right. \right. \nonumber \\
& & \hspace{1.5cm} \left. + 4 r^2 ( \omega^2 -1) \right ] + g^2_{T_2} \left [ 4 ( \omega -r)^2 - \hat{q}^2 \right ] + f^2_{T_3} \hat{q}^2 ( \omega^2 -1)^2 \nonumber \\
& & \hspace{1.5cm} \left. + 2 g_{T_1} g_{T_2} \left [ 3 \omega \hat{q}^2 + 4 r ( \omega^2 -1) \right ] - 2 g_{T_3} ( g_{T_1} \omega + g_{T_2} ) \hat{q}^2 ( \omega^2 -1) \right ) \nonumber \\
& & \hspace{1.5cm} -24 C_T \sqrt{\rho_l} \hat{q}^2 \left ( ( 1+ C_{V_L} - C_{V_R} ) 2 g_A ( g_{T_1} r + g_{T_2} ) ( \omega^2 -1) \right. \nonumber \\
& & \hspace{1.5cm} - ( 1+ C_{V_L} + C_{V_R} ) \left [ 2 g_{T_1} g_{V_1} ( 1-r \omega) + [ \omega g_{T_1} + 3 g_{T_2} - g_{T_3} ( \omega^2 -1) ] \right. \nonumber \\
& & \hspace{1.5cm} \left. \left. \times g_{V_1} ( \omega -r ) + \left [ \omega g_{T_1} + g_{T_2} - g_{T_3} ( \omega^2 -1) \right ] ( g_{V_2} r + g_{V_3} ) ( \omega^2 -1)\right ] \right ) \nonumber \\
& & \hspace{1.5cm} + 4 C_{V_R} ( 1+ C_{V_L} ) \left ( 3 g^2_{V_1} \hat{q}^2 ( 2 \hat{q}^2 + \rho_l) + 2 g_{V_1} \left [ g_{V_1} + 2 g_{V_3} ( \omega -r) \right ] ( \omega^2 -1)  \right. \nonumber \\
& & \hspace{1.5cm} \times ( \hat{q}^2 + 2 \rho_l ) + ( \omega^2 -1) \left ( g^2_{V_2} \left [ 2 r^2 \hat{q}^2 ( \omega^2 -1) + \rho_l [ 3 \hat{q}^2 + 4 r^2 ( \omega^2 -1) ] \right ] \right. \nonumber \\
& & \hspace{1.5cm} + g^2_{V_3} \left [ 2 \hat{q}^2 ( \omega^2 -1) + \rho_l [ 4 ( \omega -r)^2 - \hat{q}^2 ] \right ] + 2 g_{V_1} g_{V_2} \left [ 2 r \hat{q}^2 ( \omega -r ) + \rho_l ( 3 \right. \nonumber \\
& & \hspace{1.5cm} \left. \left. \left. \left.  - r^2 - 2 r \omega ) \right] + 2 g_{V_2} g_{V_3} \left [ 2 r \hat{q}^2 ( \omega^2 -1) + \rho_l [
3 \omega \hat{q}^2 + 4 r ( \omega^2 -1)] \right ]\right ) \right ) \right \}. \label{DWD1p}
\end{eqnarray}
 The corresponding formulae for the $B_{(s)} \rightarrow D_{(s)1} l \bar{\nu}_l$ decays can be obtained from Eqs.(\ref{DWD1pSM}), (\ref{DWD1p}) via the replacements $g_S \rightarrow -f_S$, $g_A \rightarrow f_A$, $g_{V_i} \rightarrow f_{V_i}$, $g_{T_i} \rightarrow f_{T_i}$$(i=1, 2, 3)$, and $m_{D^\prime_{(s)1}} \rightarrow m_{D_{(s)1}}$.
For the $B_{(s)} \rightarrow D^*_{(s)2} l \bar{\nu}_l$ decays,
\begin{eqnarray}
& & \frac{d \Gamma^{\rm SM}}{d q^2}= \frac{ \Gamma_0 r^3 (\omega^2-1)^{3/2}}{ 3 m_{B_{(s)}} m_{D^*_{(s)2}}}  \frac{(\hat{q}^2 - \rho_l)^2}{\hat{q}^6} \left \{ k^2_{A_1} \left [ 2 \hat{q}^2 [ 2 ( \omega -r)^2 + 3 \hat{q}^2 ] + \rho_l [ 8 ( \omega -r)^2 \right. \right. \nonumber \\
& & \hspace{1.5cm} \left. - 3 \hat{q}^2 ] \right ] + 2 ( \omega^2 -1) \left ( k^2_{A_2} \left [ 2 r^2 \hat{q}^2 ( \omega^2 -1) + \rho_l [ 3 \hat{q}^2 + 4 r^2 ( \omega^2 -1)] \right ] \right. \nonumber \\
& & \hspace{1.5cm} + k^2_{A_3} \left [ 2 \hat{q}^2 ( \omega^2 -1) + \rho_l [ 4 ( \omega -r)^2 - \hat{q}^2 ] \right ] + 3 k^2_V \hat{q}^2 ( \hat{q}^2 + \rho_l/2) \nonumber \\
& & \hspace{1.5cm} + 2 k_{A_1} k_{A_2} \left [ 2 r \hat{q}^2 ( \omega -r ) + \rho_l ( 3 -r^2 - 2 r \omega ) \right ] + 4 k_{A_1} k_{A_3} ( \omega -r ) ( \hat{q}^2 + 2 \rho_l ) \nonumber \\
& & \hspace{1.5cm} \left. \left. + 2 k_{A_2} k_{A_3} \left [ 2 r \hat{q}^2 ( \omega^2 -1) + \rho_l [ 3 \omega \hat{q}^2 + 4 r ( \omega^2 -1)] \right ]\right ) \right \}, \\
& & \frac{d \Gamma}{d q^2}=\frac{d \Gamma^{\rm SM}}{d q^2}( 1+ C_{V_L} - C_{V_R} )^2 + \frac{ 2 \Gamma_0 r^3 (\omega^2-1)^{3/2}}{ 3 m_{B_{(s)}} m_{D^*_{(s)2}}} \frac{(\hat{q}^2 - \rho_l)^2}{\hat{q}^6} \left \{ 6 C_{V_R} ( 1+ C_{V_L} ) k^2_V \right. \nonumber \\
& & \hspace{1.5cm} \times ( \omega^2 -1) \hat{q}^2 ( 2 \hat{q}^2 + \rho_l ) + 3 ( C_{S_R} - C_{S_L} )^2 k^2_P ( \omega^2 -1) \hat{q}^4 + 6 ( C_{S_L} - C_{S_R}) k_P \nonumber \\
& & \hspace{1.5cm} \times ( \omega^2 -1) \hat{q}^2 \sqrt{\rho_l} ( 1+ C_{V_L} - C_{V_R} ) \left [ k_{A_1} + k_{A_2} ( 1- r \omega ) + k_{A_3} ( \omega -r) \right ] \nonumber \\
& & \hspace{1.5cm} +16 C^2_T ( \hat{q}^2 + 2 \rho_l ) \left ( k^2_{T_1} ( \omega +1) \left [ \hat{q}^2 ( 4 \omega +1) + 6 r  ( \omega^2 -1) \right ] + k^2_{T_2} ( \omega -1 ) \right. \nonumber \\
& & \hspace{1.5cm} \times \left [ \hat{q}^2 ( 4 \omega -1) + 6r ( \omega^2 -1) \right] + k_{T_3} ( \omega^2 -1) \hat{q}^2 \left [ k_{T_3} ( \omega^2 -1) +2 k_{T_1} ( \omega +1) \right. \nonumber \\
& & \hspace{1.5cm} \left. \left. + 2 k_{T_2} ( \omega -1) \right ] - 4 k_{T_1} k_{T_2} ( \omega^2 -1) ( 1+ r \omega - 2 r^2 ) \right )
+ 12 C_T \sqrt{\rho_l} \hat{q}^2 \left ( ( \omega^2 -1) \right. \nonumber \\
& & \hspace{1.5cm} \times \left ( 2 ( k_{A_2} r + k_{A_3} ) ( 1+ C_{V_L} - C_{V_R} ) \left [ k_{T_1} ( \omega +1) + ( \omega -1) ( k_{T_2} +
k_{T_3} ( 1+ \omega ) )\right ]  \right. \nonumber \\
& & \hspace{1.5cm} \left. - 3 k_V ( 1+ C_{V_L} + C_{V_R} ) \left [ k_{T_1} ( 1+ r) - k_{T_2} ( 1-r ) \right ] \right ) + k_{A_1} ( 1+ C_{V_L} - C_{V_R} )\nonumber \\
& & \hspace{1.5cm}  \times \left[ k_{T_1} ( \omega +1) ( 3+ 2 \omega -5 r ) - k_{T_2} ( \omega -1) ( 3- 2 \omega + 5 r) \right. \nonumber \\
& & \hspace{1.5cm} \left. \left. \left. + 2 k_{T_3} ( \omega^2 -1) ( \omega -r ) \right ] \right ) \right \},
\end{eqnarray}
where
\begin{eqnarray}
\Gamma_0 = \frac{G^2_F |V_{cb}|^2 m^5_{B_{(s)}}}{192 \pi^3}, \hspace{0.5cm} \hat{q}^2= \frac{q^2}{m^2_{B_{(s)}}}, \hspace{0.5cm} \rho_l = \frac{m^2_l}{m^2_{B_{(s)}}},
\end{eqnarray}
and $m_l$ is the mass of charged lepton in the final state.

In this study, we adopt the single operator scenario, i.e. consider the contributions from the operators $O_{S_L}$, $O_{S_R}$, $O_{V_L}$, $O_{V_R}$, $O_T$ one by one and assume the corresponding Wilson coefficients to be real. Additionally, we assume that only the third generation leptons are relevant to NP for simplicity. The fitted values for the Wilson coefficients obtained in Ref.\cite{WilsonC} are as follows:
\begin{eqnarray}\label{WilsonCV}
& & C_{S_L} = 0.08 \pm 0.02, \hspace{0.5cm} C_{S_R} = -0.05 \pm 0.03, \nonumber \\
& & C_{V_L} = 0.17 \pm 0.05, \hspace{0.5cm} C_{V_R} = 0.20 \pm 0.05, \hspace{0.5cm} C_T = -0.03 \pm 0.01.
\end{eqnarray}

\begin{figure}
\centering
\includegraphics[width=3in]{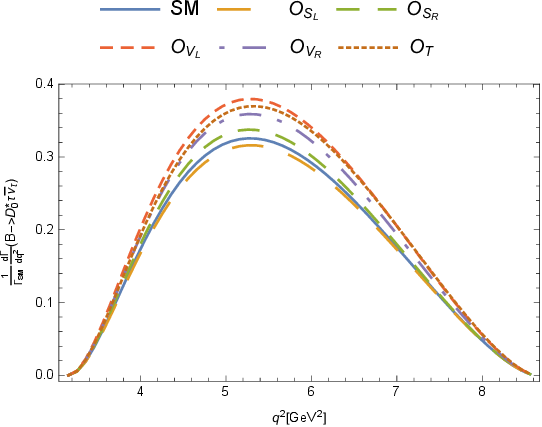} \hfill
\includegraphics[width=3in]{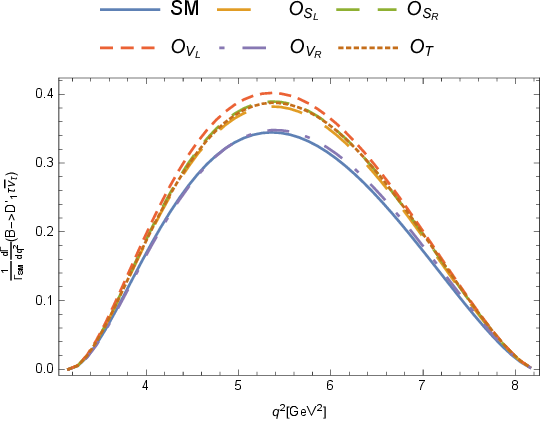} \\
\vspace{0.5cm}
\includegraphics[width=3in]{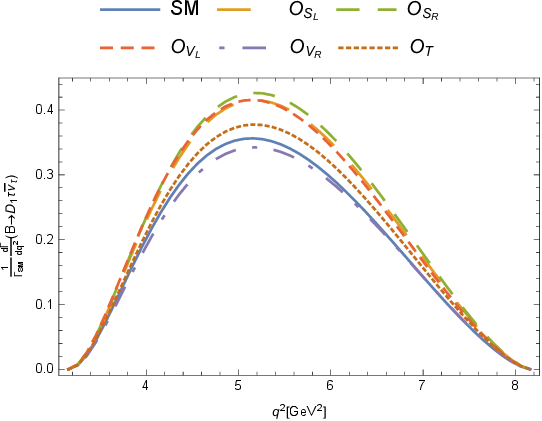} \hfill
\includegraphics[width=3in]{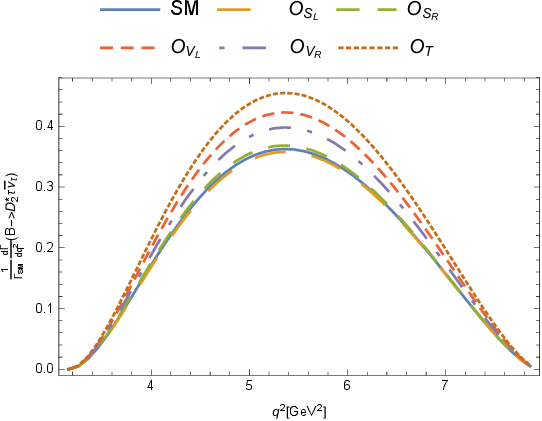} \\
\caption{Differential decay widths normalized to the SM widths for the $B \rightarrow D^{**} \tau \bar{\nu}_\tau$ decays. } \label{fig8}
\end{figure}

First, to observe the effects of each NP operators intuitively, we calculate the differential decay widths normalized to the SM widths for the $B \rightarrow D^{**} \tau \bar{\nu}_\tau$ decays, as shown in FIG.\ref{fig8}.
It is easily found that the NP effects are most significant in the moderate $q^2$ region, around $q^2 \in [4.5, 6.5]{\rm GeV^2}$. For the $B \rightarrow D^*_0 \tau \bar{\nu}_\tau$ decay, all operators except $O_{S_R}$ give positive contributions to the differential decay width, and the operator $O_{V_L}$ has a maximal contribution.
The operator $O_{V_L}$ also has a maximal contribution, and the contribution of $O_{V_R}$ is almost zero for the $B \rightarrow D^\prime_1 \tau \bar{\nu}_\tau$ decay. In contrast, the operators $O_{S_R}$ and $O_T$ have maximal contributions for the $B \rightarrow D_1 \tau \bar{\nu}_\tau$ and $B \rightarrow D^*_2 \tau \bar{\nu}_\tau$ decays, respectively. For the former decay, only the operator $O_{V_R}$ has a negative contribution to the differential decay width, and the contributions of $O_{S_L}$ and $O_{S_R}$ are nearly zero for the later decay.
For the $B_s \rightarrow D^{**}_s \tau \bar{\nu}_\tau$ decays, the corresponding behaviors of the differential decay widths normalized to the SM widths are similar.

Integrating the differential decay widths over $q^2$ in the entire physical region and using the lifetimes of $B_{(s)}$ as inputs, we can obtain the branching fractions
\begin{eqnarray}
Br = \frac{\tau_{B_{(s)}}}{\hbar} \Gamma = \frac{\tau_{B_{(s)}}}{\hbar} \int^{( m_{B_{(s)}} - m_{D^{**}_{(s)}})^2}_{m^2_l} d q^2 \frac{d \Gamma}{d q^2}.
\end{eqnarray}
Similar to Eq.(\ref{RDDstar}), the corresponding ratios of branching fractions
\begin{eqnarray}
R(D^{**}_{(s)}) = \frac{Br  (B_{(s)} \rightarrow D^{**}_{(s)} \tau \bar{\nu}_\tau  )}{Br (B_{(s)} \rightarrow D^{**}_{(s)} l \bar{\nu}_l  )} =
\frac{\Gamma  (B_{(s)} \rightarrow D^{**}_{(s)} \tau \bar{\nu}_\tau  )}{\Gamma (B_{(s)} \rightarrow D^{**}_{(s)} l \bar{\nu}_l  )}, \hspace{1cm} l =e, \mu.
\end{eqnarray}

For the lifetimes of $B_{(s)}$, masses of charged leptons, CKM matrix elements $|V_{cb}|$, Fermi coupling constant $G_F$, and reduced Plank constant $\hbar$, we use the latest values given by the PDG\cite{PDG2022},
\begin{eqnarray}
& & \tau_{B^-} = 1.638 \times 10^{-12} {\rm s}, \hspace{0.5cm} \tau_{B_s} = 1.516 \times 10^{-12} {\rm s}, \hspace{0.5cm} |V_{cb}| = (40.8 \pm 1.4) \times 10^{-3},
\nonumber \\
& & m_e= 0.51 \times 10^{-3} {\rm GeV}, \hspace{0.5cm} m_\mu = 0.106 {\rm GeV}, \hspace{0.5cm} m_\tau = 1.777 {\rm GeV}, \nonumber \\
& & G_F= 1.166 \times 10^{-5} {\rm GeV}^{-2}, \hspace{0.5cm} \hbar = 6.582 \times 10^{-25} {\rm GeV} \cdot {\rm s}.
\end{eqnarray}

The numerical results of $Br$ and $R(D^{**}_{(s)})$ for the $B_{(s)} \rightarrow D^{**}_{(s)} l \bar{\nu}_l$ decays are shown in TABLE~\ref{table5}-\ref{table8}. For specificity, we give the SM results of these two observables for the $B^- \rightarrow D^{**0} l^- \bar{\nu}_l$ and $\bar{B}^0_s \rightarrow D^{**+}_s l^- \bar{\nu}_l$ decays in TABLE~\ref{table5} and \ref{table6}, respectively. For comparison, the values given by current experiments\cite{HFLAV,PDG2022} and several previous theoretical analyses\cite{HQETE4,HQETE5,HQEFTSR1,HQEFTSR2} are also listed.
(Note that the original values from Ref.\cite{HFLAV,PDG2022} are modified using the expected absolute $D^{**}$ decay branching fractions, as detailed in Ref.\cite{HQETE5}, and the theoretical results of $Br$ for $\bar{B}^0$ decays in Ref.\cite{HQETE5,HQEFTSR2} are converted to the case of $B^-$ here.) $Br$ are at the order of ${\cal O} (10^{-4})$ and ${\cal O} ( 10^{-3})$ for the decays with the final states $D^{1/2+}_{(s)}+e\bar{\nu}_e(\mu \bar{\nu}_\mu)$ and $D^{3/2+}_{(s)}+e\bar{\nu}_e(\mu \bar{\nu}_\mu)$, respectively, and the corresponding results for the decays with $\tau$ in the final states are smaller by more than one order of magnitude. $R(D^{**}_{(s)})$ are almost identical for $l=e, \mu$, and thus we only consider the case of $l=\mu$ for this quantity when investigating the NP effects. Overall, our results are compatible with the corresponding values given by current experiments and previous theoretical analyses. Note that there exist large deviations between experimental values and theoretical predictions for the branching fractions of the $B \rightarrow D^{1/2+} l \bar{\nu}_l$ decays, which is just the `1/2 vs 3/2 puzzle'. For this puzzle, several explanations have been proposed, such as a virtual $D^{(*)}_V$ in the broad structure, relativistic corrections, and mixing between $^1P_1$ and $^3P_1$ partial waves in $D^\prime_1$\cite{puzzle2,HQETE5}.
The first and second uncertainties of $Br$ originate from the form factors and CKM matrix element $|V_{cb}|$, respectively. The former are approximately $45\%-50\%$ for all the relevant decays except the modes with $D^*_{(s)2}$ in the final states (around $70\%$ for these modes), and the later are about $7\%$. In contrast, the uncertainties of $R(D^{**}_{(s)})$ stem from the form factors and are considerably smaller because most uncertainties are canceled in this observable.

The results of $Br$ and $R(D^{**}_{(s)})$ for the $B_{(s)} \rightarrow D^{**}_{(s)} \tau \bar{\nu}_\tau$ decays with the contributions of $O_{S_L}$, $O_{S_R}$ and $O_{V_L}$, $O_{V_R}$, $O_T$ are shown in TABLE~\ref{table7} and \ref{table8}, respectively. The additional uncertainties arise from the corresponding Wilson coefficients $C_X$$(X=S_L, S_R, V_L, V_R, T)$, c.f. Eq.(\ref{WilsonCV}), which are lower than $8\%$.
The operators $O_{V_L}$, $O_{S_R}$, and $O_T$ have maximal contributions to the decays with $D^*_{(s)0}(D^\prime_{(s)1}$), $D_{(s)1}$, and $D^*_{(s)2}$ in the final states, respectively, increasing the $Br$ and $R(D^{**}_{(s)})$ for the corresponding decay modes by roughly $17\%$, $20\%$, and $25\%$.
In addition, the operators $O_{S_L}$ and $O_{V_R}$ have negative contributions to the $B_{(s)} \rightarrow D^*_{(s)0} \tau \bar{\nu}_\tau$ and
$B_{(s)} \rightarrow D_{(s)1} \tau \bar{\nu}_\tau$ decays, respectively, the effects of which are insignificant.
\begin{table}
\centering
\begin{tabular}{|c|c|c|c|c|}
\hline Decays & Obs. & $e\bar{\nu}_e$ & $\mu \bar{\nu}_\mu$ & $\tau \bar{\nu}_\tau$  \\
\hline
$B^- \rightarrow D^{*0}_0 l^- \bar{\nu}_l $ & $Br \times 10^4 $ & $6.11^{+2.79+0.36}_{-2.12-0.47}$ &  $6.05^{+2.76+0.36}_{-2.09-0.46}$ & $0.38^{+0.17+0.02}_{-0.13-0.03}$  \\
\cline{3-5}
& & $42.1 \pm 7.5$\cite{HFLAV} & $42.1 \pm 7.5$\cite{HFLAV} & ---\\
\cline{3-5}
& & $5.50 \pm 1.29$\cite{HQETE5} & $5.50 \pm 1.29$\cite{HQETE5} & $0.54 \pm 0.14$\cite{HQETE5} \\
\cline{3-5}
& & $5.31 \pm 1.70$\cite{HQEFTSR2} & $5.31 \pm 1.70$\cite{HQEFTSR2} & ---\\
\cline{2-5}
& $R (D^*_0)$ & $0.063^{+0.001}_{-0.002}$ &  $0.063^{+0.001}_{-0.002}$  &  ---\\
\cline{3-5}
& & $0.08 \pm 0.03$\cite{HQETE4} & $0.08 \pm 0.03$\cite{HQETE4} & ---\\
\cline{3-5}
& & $0.099 \pm 0.015$\cite{HQETE5} & $0.099 \pm 0.015$\cite{HQETE5} & ---\\
\hline
$B^- \rightarrow D^{\prime 0}_1 l^- \bar{\nu}_l $ & $Br \times 10^4$ & $6.76^{+3.09+0.40}_{-2.35-0.52}$ &  $6.70^{+3.06+0.40}_{-2.33-0.51}$ & $0.52^{+0.24+0.03}_{-0.18-0.04}$  \\
\cline{3-5}
& & $19.4 \pm 16.2$\cite{HFLAV} & $19.4 \pm 16.2$\cite{HFLAV} & ---\\
\cline{3-5}
& & $4.96 \pm 3.99$\cite{HQETE5} & $4.96 \pm 3.99$\cite{HQETE5} & $0.37 \pm 0.29$\cite{HQETE5} \\
\cline{3-5}
& & $5.10 \pm 1.70$\cite{HQEFTSR2} & $5.10 \pm 1.70$\cite{HQEFTSR2} & ---\\
\cline{2-5}
& $R (D^\prime_1)$ & $0.076^{+0.001}_{-0.002}$ &  $0.077^{+0.001}_{-0.002}$  &  ---\\
\cline{3-5}
& & $0.05 \pm 0.02$\cite{HQETE4} & $0.05 \pm 0.02$\cite{HQETE4} & ---\\
\cline{3-5}
& & $0.074 \pm 0.012$\cite{HQETE5} & $0.074 \pm 0.012$\cite{HQETE5} & ---\\
\hline
$B^- \rightarrow D^0_1 l^- \bar{\nu}_l $ & $Br \times 10^3 $ & $7.26^{+3.60+0.51}_{-2.87-0.49}$ &  $7.19^{+3.56+0.50}_{-2.84-0.48}$ & $0.46^{+0.22+0.03}_{-0.17-0.03}$  \\
\cline{3-5}
& & $6.73 \pm 0.58$\cite{PDG2022} & $6.73 \pm 0.58$\cite{PDG2022} & ---\\
\cline{3-5}
& & $6.90 \pm 0.47$\cite{HQETE5} & $6.90 \pm 0.47$\cite{HQETE5} & $0.68 \pm 0.06$\cite{HQETE5} \\
\cline{3-5}
& & $5.3 \pm 1.6$\cite{HQEFTSR1} & $5.3 \pm 1.6$\cite{HQEFTSR1} & ---\\
\cline{2-5}
& $R(D_1)$ & $0.064^{+0.003}_{-0.003}$ &  $0.064^{+0.004}_{-0.002}$  &  ---\\
\cline{3-5}
& & $0.10 \pm 0.02$\cite{HQETE4} & $0.10 \pm 0.02$\cite{HQETE4} & ---\\
\cline{3-5}
& & $0.098 \pm 0.007$\cite{HQETE5} & $0.098 \pm 0.007$\cite{HQETE5} & ---\\
\hline
$B^- \rightarrow D^{* 0}_2 l^- \bar{\nu}_l $ & $Br \times 10^3$ & $3.65^{+2.67+0.25}_{-2.01-0.25}$ &  $3.61^{+2.65+0.25}_{-1.99-0.24}$ & $0.25^{+0.17+0.02}_{-0.13-0.02}$  \\
\cline{3-5}
& & $3.43 \pm 0.28$\cite{PDG2022} & $3.43 \pm 0.28$\cite{PDG2022} & ---\\
\cline{3-5}
& & $3.40 \pm 0.32$\cite{HQETE5} & $3.40 \pm 0.32$\cite{HQETE5} & $0.20 \pm 0.03$\cite{HQETE5}\\
\cline{3-5}
& & $4.0 \pm 2.6$\cite{HQEFTSR1} & $4.0 \pm 2.6$\cite{HQEFTSR1} & ---\\
\cline{2-5}
& $R (D^*_2)$ & $0.068^{+0.003}_{-0.002}$ &  $0.069^{+0.002}_{-0.003}$  &  ---\\
\cline{3-5}
& & $0.07 \pm 0.01$\cite{HQETE4} & $0.07 \pm 0.01$\cite{HQETE4} & ---\\
\cline{3-5}
& & $0.060 \pm 0.005$\cite{HQETE5} & $0.060 \pm 0.005$\cite{HQETE5} & ---\\
\hline
\end{tabular}
\caption{Numerical results of $Br$ and $R(D^{**})$ in the SM for $B \rightarrow D^{**} l \bar{\nu}_l$ decays.}\label{table5}
\end{table}
\begin{table}
\centering
\begin{tabular}{|c|c|c|c|c|}
\hline Decays & Obs. & $e\bar{\nu}_e$ & $\mu \bar{\nu}_\mu$ & $\tau \bar{\nu}_\tau$  \\
\hline
$\bar{B}^0_s \rightarrow D^{*+}_{s0} l^- \bar{\nu}_l $ & $Br \times 10^4$ & $7.18^{+3.34+0.43}_{-2.55-0.55}$ &  $7.11^{+3.31+0.42}_{-2.53-0.54}$ & $0.57^{+0.27+0.03}_{-0.21-0.04}$  \\
\cline{2-5}
& $R (D^*_{s0})$ & $0.080^{+0.001}_{-0.002}$ &  $0.080^{+0.001}_{-0.002}$  &  ---\\
\cline{3-5}
& & $0.09 \pm 0.04$\cite{HQETE3} & $0.09 \pm 0.04$\cite{HQETE3} & ---\\
\hline
$\bar{B}^0_s \rightarrow D^{\prime +}_{s1} l^- \bar{\nu}_l $ & $Br \times 10^4$ & $6.48^{+3.03+0.38}_{-2.32-0.50}$ &  $6.42^{+3.00+0.38}_{-2.30-0.49}$ & $0.54^{+0.25+0.03}_{-0.20-0.04}$  \\
\cline{2-5}
& $R(D^\prime_{s1})$ & $0.084^{+0.001}_{-0.002}$ &  $0.084^{+0.001}_{-0.002}$  &  ---\\
\cline{3-5}
& & $0.07 \pm 0.03$\cite{HQETE3} & $0.07 \pm 0.03$\cite{HQETE3} & ---\\
\hline
$\bar{B}^0_s \rightarrow D^+_{s1} l^- \bar{\nu}_l $ & $Br \times 10^3$ & $6.31^{+3.07+0.44}_{-2.44-0.43}$ &  $6.25^{+3.03+0.44}_{-2.42-0.42}$ & $0.38^{+0.18+0.03}_{-0.14-0.03}$  \\
\cline{2-5}
& $R(D_{s1})$ & $0.061^{+0.003}_{-0.003}$ &  $0.061^{+0.003}_{-0.002}$  &  ---\\
\cline{3-5}
& & $0.09 \pm 0.02$\cite{HQETE3} & $0.09 \pm 0.02$\cite{HQETE3} & ---\\
\hline
$\bar{B}^0_s \rightarrow D^{*+}_{s2} l^- \bar{\nu}_l $ & $Br \times 10^3$ & $3.77^{+2.59+0.26}_{-1.95-0.25}$ &  $3.73^{+2.56+0.26}_{-1.93-0.25}$ & $0.24^{+0.16+0.02}_{-0.12-0.02}$  \\
\cline{2-5}
& $R (D^*_{s2})$ & $0.063^{+0.003}_{-0.001}$ &  $0.064^{+0.002}_{-0.002}$  &  ---\\
\cline{3-5}
& & $0.07 \pm 0.01$\cite{HQETE3} & $0.07 \pm 0.01$\cite{HQETE3} & ---\\
\hline
\end{tabular}
\caption{Numerical results of $Br$ and $R(D^{**}_s)$ in the SM for $B_s \rightarrow D^{**}_s l \bar{\nu}_l$ decays.  }\label{table6}
\end{table}
\begin{table}
\centering
\begin{tabular}{|c|c|c|c|}
\hline Decays & Obs. & $O_{S_L}$ & $O_{S_R}$  \\
\hline
$B^- \rightarrow D^{*0}_0 \tau^- \bar{\nu}_\tau $ & $Br \times 10^4$ & $0.37^{+0.17+0.02+0.00}_{-0.13-0.03-0.00} $ &  $0.40^{+0.18+0.02+0.00}_{-0.14-0.03-0.00}$ \\
\cline{2-4}
& $R (D^*_0)$ & $0.062^{+0.001+0.000}_{-0.002-0.000}$ &  $0.065^{+0.001+0.001}_{-0.002-0.001}$   \\
\hline
$\bar{B}^0_s \rightarrow D^{*+}_{s0} \tau^- \bar{\nu}_\tau $ & $Br \times 10^4$ & $0.56^{+0.26+0.03+0.00}_{-0.20-0.04-0.00} $ &  $0.59^{+0.28+0.04+0.01}_{-0.21-0.05-0.01}$ \\
\cline{2-4}
& $R (D^*_{s0})$ & $0.078^{+0.001+0.001}_{-0.002-0.001}$ &  $0.083^{+0.002+0.001}_{-0.002-0.001}$   \\
\hline
$B^- \rightarrow D^{\prime 0}_1 \tau^- \bar{\nu}_\tau $ & $Br \times 10^4$ & $0.57^{+0.26+0.03+0.02}_{-0.20-0.04-0.02} $ &  $0.58^{+0.27+0.03+0.02}_{-0.20-0.04-0.02}$ \\
\cline{2-4}
& $R(D^\prime_1)$ & $0.085^{+0.002+0.003}_{-0.002-0.003}$ &  $0.087^{+0.002+0.003}_{-0.002-0.003}$   \\
\hline
$\bar{B}^0_s \rightarrow D^{\prime +}_{s1} \tau^- \bar{\nu}_\tau $ & $Br \times 10^4$ & $0.60^{+0.28+0.04+0.02}_{-0.22-0.05-0.02}$ &  $0.61^{+0.29+0.04+0.02}_{-0.22-0.05-0.02}$ \\
\cline{2-4}
& $R(D^\prime_{s1})$ & $0.093^{+0.001+0.003}_{-0.002-0.003}$ &  $0.095^{+0.002+0.003}_{-0.002-0.003}$ \\
\hline
$B^- \rightarrow D^0_1 \tau^- \bar{\nu}_\tau $ & $Br \times 10^3$ & $0.54^{+0.26+0.04+0.03}_{-0.22-0.04-0.02} $ &  $0.56^{+0.28+0.04+0.03}_{-0.22-0.04-0.03}$ \\
\cline{2-4}
& $R (D_1)$ & $0.075^{+0.003+0.004}_{-0.002-0.003}$ &  $0.077^{+0.003+0.004}_{-0.001-0.004}$   \\
\hline
$\bar{B}^0_s \rightarrow D^+_{s1} \tau^- \bar{\nu}_\tau $ & $Br \times 10^3$ & $0.45^{+0.22+0.03+0.02}_{-0.17-0.03-0.02} $ &  $0.46^{+0.23+0.03+0.02}_{-0.18-0.03-0.02}$ \\
\cline{2-4}
& $R (D_{s1})$ & $0.072^{+0.002+0.003}_{-0.002-0.003}$ &  $0.074^{+0.002+0.004}_{-0.002-0.003}$   \\
\hline
$B^- \rightarrow D^{*0}_2 \tau^- \bar{\nu}_\tau $ & $Br \times 10^3$ & $0.25^{+0.17+0.02+0.00}_{-0.14-0.02-0.00} $ &  $0.25^{+0.18+0.02+0.00}_{-0.13-0.02-0.00}$ \\
\cline{2-4}
& $R(D^*_2)$ & $0.068^{+0.002+0.000}_{-0.003-0.000}$ &  $0.070^{+0.002+0.000}_{-0.003-0.000}$   \\
\hline
$\bar{B}^0_s \rightarrow D^{*+}_{s2} \tau^- \bar{\nu}_\tau $ & $Br \times 10^3$ & $0.24^{+0.16+0.02+0.00}_{-0.12-0.02-0.00} $ &  $0.24^{+0.16+0.02+0.00}_{-0.12-0.02-0.00}$ \\
\cline{2-4}
& $R(D^*_{s2})$ & $0.063^{+0.002+0.000}_{-0.001-0.000}$ &  $0.065^{+0.002+0.000}_{-0.002-0.000}$   \\
\hline
\end{tabular}
\caption{Numerical results of $Br$ and $R(D^{**}_{(s)})$ for $B_{(s)} \rightarrow D^{**}_{(s)} \tau \bar{\nu}_\tau$ decays with the contributions of $O_{S_L}$ and $O_{S_R}$.   }\label{table7}
\end{table}
\begin{table}
\centering
\begin{tabular}{|c|c|c|c|c|}
\hline Decays & Obs. & $O_{V_L}$ & $O_{V_R}$ & $O_T$ \\
\hline
$B^- \rightarrow D^{*0}_0 \tau^- \bar{\nu}_\tau $ & $Br \times 10^4$ & $0.45^{+0.20+0.03+0.02}_{-0.16-0.03-0.02}$ &  $0.42^{+0.19+0.03+0.02}_{-0.15-0.03-0.02}$ & $0.44^{+0.20+0.03+0.02}_{-0.15-0.03-0.02}$\\
\cline{2-5}
& $R(D^*_0)$ & $0.074^{+0.002+0.003}_{-0.002-0.003}$ &  $0.070^{+0.001+0.004}_{-0.002-0.004}$ & $0.072^{+0.002+0.003}_{-0.002-0.003}$   \\
\hline
$\bar{B}^0_s \rightarrow D^{*+}_{s0} \tau^- \bar{\nu}_\tau $ & $Br \times 10^4$ & $0.67^{+0.31+0.04+0.02}_{-0.24-0.05-0.02}$ &  $0.63^{+0.29+0.04+0.04}_{-0.23-0.05-0.04}$ & $0.65^{+0.30+0.04+0.03}_{-0.23-0.05-0.03}$\\
\cline{2-5}
& $R (D^*_{s0})$ & $0.094^{+0.002+0.004}_{-0.003-0.003}$ &  $0.089^{+0.002+0.005}_{-0.003-0.005}$ & $0.091^{+0.002+0.004}_{-0.003-0.004}$   \\
\hline
$B^- \rightarrow D^{\prime 0}_1 \tau^- \bar{\nu}_\tau $ & $Br \times 10^4$ & $0.60^{+0.27+0.04+0.02}_{-0.21-0.05-0.02}$ &  $0.52^{+0.24+0.03+0.01}_{-0.18-0.04-0.00}$ & $0.58^{+0.26+0.03+0.02}_{-0.20-0.04-0.02}$\\
\cline{2-5}
& $R (D^\prime_1)$ & $0.090^{+0.002+0.003}_{-0.003-0.003}$ &  $0.078^{+0.001+0.001}_{-0.002-0.001}$ & $0.087^{+0.002+0.004}_{-0.002-0.003}$   \\
\hline
$\bar{B}^0_s \rightarrow D^{\prime +}_{s1} \tau^- \bar{\nu}_\tau $ & $Br \times 10^4$ & $0.63^{+0.30+0.04+0.02}_{-0.23-0.05-0.02}$ &  $0.55^{+0.26+0.03+0.01}_{-0.20-0.04-0.00}$ & $0.61^{+0.29+0.04+0.03}_{-0.22-0.05-0.02}$\\
\cline{2-5}
& $R(D^\prime_{s1})$ & $0.098^{+0.002+0.004}_{-0.002-0.004}$ &  $0.085^{+0.001+0.001}_{-0.002-0.001}$ & $0.095^{+0.002+0.004}_{-0.002-0.004}$   \\
\hline
$B^- \rightarrow D^0_1 \tau^- \bar{\nu}_\tau $ & $Br \times 10^3$ & $0.54^{+0.26+0.04+0.02}_{-0.20-0.04-0.02}$ &  $0.45^{+0.20+0.03+0.01}_{-0.16-0.03-0.01}$ & $0.49^{+0.23+0.03+0.01}_{-0.18-0.03-0.01}$\\
\cline{2-5}
& $R(D_1)$ & $0.075^{+0.004+0.003}_{-0.003-0.003}$ &  $0.062^{+0.005+0.001}_{-0.003-0.001}$ & $0.069^{+0.006+0.002}_{-0.005-0.002}$   \\
\hline
$\bar{B}^0_s \rightarrow D^+_{s1} \tau^- \bar{\nu}_\tau $ & $Br \times 10^3$ & $0.45^{+0.21+0.03+0.02}_{-0.17-0.03-0.02}$ &  $0.37^{+0.18+0.03+0.01}_{-0.13-0.02-0.01}$ & $0.41^{+0.18+0.03+0.01}_{-0.15-0.03-0.01}$\\
\cline{2-5}
& $R(D_{s1})$ & $0.071^{+0.004+0.003}_{-0.002-0.003}$ &  $0.059^{+0.004+0.001}_{-0.003-0.001}$ & $0.065^{+0.005+0.002}_{-0.003-0.001}$   \\
\hline
$B^- \rightarrow D^{*0}_2 \tau^- \bar{\nu}_\tau $ & $Br \times 10^3$ & $0.29^{+0.20+0.02+0.01}_{-0.16-0.02-0.01}$ &  $0.27^{+0.19+0.02+0.02}_{-0.14-0.02-0.01}$ & $0.31^{+0.21+0.02+0.02}_{-0.16-0.02-0.02}$\\
\cline{2-5}
& $R(D^*_2)$ & $0.080^{+0.004+0.003}_{-0.002-0.003}$ &  $0.075^{+0.003+0.004}_{-0.002-0.004}$ & $0.086^{+0.005+0.007}_{-0.004-0.006}$   \\
\hline
$\bar{B}^0_s \rightarrow D^{*+}_{s2} \tau^- \bar{\nu}_\tau $ & $Br \times 10^3$ & $0.28^{+0.18+0.02+0.01}_{-0.14-0.02-0.01}$ &  $0.26^{+0.17+0.02+0.01}_{-0.13-0.02-0.01}$ & $0.30^{+0.19+0.02+0.02}_{-0.14-0.02-0.02}$\\
\cline{2-5}
& $R(D^*_{s2})$ & $0.075^{+0.002+0.003}_{-0.003-0.003}$ &  $0.070^{+0.003+0.004}_{-0.001-0.004}$ & $0.079^{+0.005+0.006}_{-0.002-0.006}$   \\
\hline
\end{tabular}
\caption{Numerical results of $Br$ and $R(D^{**}_{(s)})$ for $B_{(s)} \rightarrow D^{**}_{(s)} \tau \bar{\nu}_\tau$ decays with the contributions of $O_{V_L}$, $O_{V_R}$ and $O_T$.}\label{table8}
\end{table}

\section{Summary}

In this study, we calculate the $B_{(s)} \rightarrow D^{**}_{(s)}$ form factors systematically using QCD sum rules in the framework of HQEFT and perform a model independent analysis of the corresponding semileptonic decays, including the contributions from possible NP effects.
We consider contributions up to the next leading order of heavy quark expansion and give all the relevant form factors, including the scalar and tensor ones only related to the NP effects. Expressions for the form factors in terms of universal wave functions are derived via heavy quark expansion, and several relations among the form factors are obtained, i.e. $g_- = g_T$, $g_{V_2} = - g_{T_3}$, $g_A=-g_{T_1} = g_{T_2}$, $f_A = - f_{T_1}= f_{T_2}$, and $k_{A_2} = k_{T_3}$. We find that the form factor $g_+$ is equal to zero in the entire physical region of $q^2$, and $g_P$, $g_{V_1}$, $f_{V_1}$ approach zero at $q^2= q^2_{max}$. Neglecting the contributions from chromomagnetic operators for the $B_{(s)} \rightarrow D^{1/2+}_{(s)}$ decays, we have the additional relations $g_{V_2}=-g_{T_3}=0$ and $g_S =-g_{V_3}=g_A$. The values of the $B \rightarrow D^{**}$ form factors are very close to their strange counterparts owing to the approximate $SU(3)$ flavor symmetry. For the $B_{(s)} \rightarrow D^{1/2+}_{(s)}$ decays, the uncertainties of the form factors are approximately $20\%-30\%$, while the maximum uncertainties can reach $90\%$ for the decays with $D^{3/2+}_{(s)}$ in the final states.

With the form factors given here, we analyze the relevant semileptonic decays $B_{(s)} \rightarrow D^{**}_{(s)} l \bar{\nu}_l$ model independently, including the NP contributions from possible (pseudo-)scalar, (axial-)vector, and tensor interactions. We assume the active neutrinos to be left-handed and consider these contributions in the single operator scenario. In addition, the Wilson coefficients are assumed to be real and the NP only relevant to the third generation leptons, for simplicity. It is found that the branching fractions are at the order of ${\cal O}(10^{-4})$ and ${\cal O}(10^{-3})$ for the decays with the final states $D^{1/2+}_{(s)}+e\bar{\nu}_e(\mu \bar{\nu}_\mu)$ and $D^{3/2+}_{(s)}+e\bar{\nu}_e(\mu \bar{\nu}_\mu)$, respectively, and the corresponding results for the decays with $\tau$ in the final states are smaller by more than one order of magnitude. The NP effects are most significant in the moderate $q^2$ region, i.e. around $q^2 \in [4.5, 6.5]{\rm GeV}^2$.
The operators $O_{V_L}$, $O_{S_R}$, and $O_T$ have maximal contributions to the decays with $D^*_{(s)0}(D^\prime_{(s)1}$), $D_{(s)1}$, and $D^*_{(s)2}$ in the final states, respectively, increasing the branching fractions and the ratios $R(D^{**}_{(s)})$ for the corresponding decay modes by roughly $17\%$, $20\%$, and $25\%$, respectively. In contrast, the operators $O_{S_L}$ and $O_{V_R}$ have insignificant negative contributions to the $B_{(s)} \rightarrow D^*_{(s)0} \tau \bar{\nu}_\tau$ and
$B_{(s)} \rightarrow D_{(s)1} \tau \bar{\nu}_\tau$ decays, respectively. Our results may be tested in more precise experiments in the future.




\appendix

 \section{The extraction of $\kappa_1(1)$, $\kappa_2(1)$, $\kappa^{1/2}_1(1)$, $\kappa^{1/2}_2(1)$, $\kappa^{3/2}_1(1)$, $\kappa^{3/2}_2(1)$ and $\bar{\Lambda}_{(s)}$, $\bar{\Lambda}^{1/2}_{(s)}$, $\bar{\Lambda}^{3/2}_{(s)}$ by fitting the meson masses}\label{kappaPs}

According to Ref.\cite{HQEFT3,HQEFTSR1,HQEFTSR2}, the Lorentz scalar factors $\kappa_i(\omega)$, $\kappa^{1/2}_i(\omega)$, $\kappa^{3/2}_i(\omega)$ $(i=1,2)$ are defined
in HQEFT via the next leading order hadronic matrix elements of $j^P_l = \frac{1}{2}^-, \frac{1}{2}^+, \frac{3}{2}^+$ heavy mesons, respectively.
Now we show that their values at zero recoiling point ($\omega =1$) can be extracted by fitting the meson masses.

To the next leading order of heavy quark expansion, the binding energies of the $j^P_l = \frac{1}{2}^-$ heavy mesons can be written as\cite{HQEFTSR1,HQEFTSR2}
\begin{eqnarray}
& & \bar{\Lambda}_{B_{(s)}} = \bar{\Lambda}_{(s)} - \frac{1}{m_b} \left ( \kappa_1(1) + 3 \kappa_2(1) \right ), \label{B}\\
& & \bar{\Lambda}_{B^*_{(s)}} = \bar{\Lambda}_{(s)} - \frac{1}{m_b} \left ( \kappa_1(1) -  \kappa_2(1) \right ),\label{B*}\\
& & \bar{\Lambda}_{D_{(s)}} = \bar{\Lambda}_{(s)} - \frac{1}{m_c} \left ( \kappa_1(1) + 3 \kappa_2(1) \right ), \label{D}\\
& & \bar{\Lambda}_{D^*_{(s)}} = \bar{\Lambda}_{(s)} - \frac{1}{m_c} \left ( \kappa_1(1) -  \kappa_2(1) \right ), \label{D*}
\end{eqnarray}
where
\begin{eqnarray}
& & \bar{\Lambda}_{B_{(s)}} = m_{B_{(s)}} - m_b, \hspace{0.5cm} \bar{\Lambda}_{B^*_{(s)}} = m_{B^*_{(s)}} - m_b, \label{BB*}\\
& & \bar{\Lambda}_{D_{(s)}} = m_{D_{(s)}} - m_c, \hspace{0.5cm} \bar{\Lambda}_{D^*_{(s)}} = m_{D^*_{(s)}} - m_c. \label{DD*}
\end{eqnarray}
For the $j^P_l = \frac{1}{2}^+$ heavy mesons, the binding energies are
\begin{eqnarray}
& & \bar{\Lambda}_{B^*_{(s)0}} = \bar{\Lambda}_{(s)} - \frac{1}{m_b} \left ( \kappa^{1/2}_1(1) + 3 \kappa^{1/2}_2(1) \right ), \label{B0}\\
& & \bar{\Lambda}_{B^\prime_{(s)1}} = \bar{\Lambda}_{(s)} - \frac{1}{m_b} \left ( \kappa^{1/2}_1(1) -  \kappa^{1/2}_2(1) \right ), \label{B1p}\\
& & \bar{\Lambda}_{D^*_{(s)0}} = \bar{\Lambda}_{(s)} - \frac{1}{m_c} \left ( \kappa^{1/2}_1(1) + 3 \kappa^{1/2}_2(1) \right ), \label{D0}\\
& & \bar{\Lambda}_{D^\prime_{(s)1}} = \bar{\Lambda}_{(s)} - \frac{1}{m_c} \left ( \kappa^{1/2}_1(1) -  \kappa^{1/2}_2(1) \right ), \label{D1p}
\end{eqnarray}
where
\begin{eqnarray}
& & \bar{\Lambda}_{B^*_{(s)0}} = m_{B^*_{(s)0}} - m_b, \hspace{0.5cm} \bar{\Lambda}_{B^\prime_{(s)1}} = m_{B^\prime_{(s)1}} - m_b, \label{B0B1p}\\
& & \bar{\Lambda}_{D^*_{(s)0}} = m_{D^*_{(s)0}} - m_c, \hspace{0.5cm} \bar{\Lambda}_{D^\prime_{(s)1}} = m_{D^\prime_{(s)1}} - m_c. \label{D0D1p}
\end{eqnarray}
For the $j^P_l = \frac{3}{2}^+$ heavy mesons,
\begin{eqnarray}
& & \bar{\Lambda}_{B_{(s)1}} = \bar{\Lambda}_{(s)} - \frac{1}{m_b} \left ( \kappa^{3/2}_1(1) + 5 \kappa^{3/2}_2(1) \right ), \label{B1}\\
& & \bar{\Lambda}_{B^*_{(s)2}} = \bar{\Lambda}_{(s)} - \frac{1}{m_b} \left ( \kappa^{3/2}_1(1) - 3 \kappa^{3/2}_2(1) \right ), \label{B2}\\
& & \bar{\Lambda}_{D_{(s)1}} = \bar{\Lambda}_{(s)} - \frac{1}{m_c} \left ( \kappa^{3/2}_1(1) + 5 \kappa^{3/2}_2(1) \right ), \label{D1}\\
& & \bar{\Lambda}_{D^*_{(s)2}} = \bar{\Lambda}_{(s)} - \frac{1}{m_c} \left ( \kappa^{3/2}_1(1) - 3 \kappa^{3/2}_2(1) \right ),\label{D2}
\end{eqnarray}
where
\begin{eqnarray}
& & \bar{\Lambda}_{B_{(s)1}} = m_{B_{(s)1}} - m_b, \hspace{0.5cm} \bar{\Lambda}_{B^*_{(s)2}} = m_{B^*_{(s)2}} - m_b, \label{B1B2}\\
& & \bar{\Lambda}_{D_{(s)1}} = m_{D_{(s)1}} - m_c, \hspace{0.5cm} \bar{\Lambda}_{D^*_{(s)2}} = m_{D^*_{(s)2}} - m_c. \label{D1D2}
\end{eqnarray}

Combining Eqs.(\ref{B})-(\ref{DD*}), we obtain
\begin{eqnarray}
& & \kappa_1(1)= \frac{m_b m_c}{m_b - m_c} ( \bar{m}_{B_{(s)}} - \bar{m}_{D_{(s)}} -m_b +m_c), \\
& & \kappa_2(1) = \frac{1}{4}m_c ( m_{D^*_{(s)}}-m_{D_{(s)}}),
\end{eqnarray}
and
\begin{eqnarray}
\bar{\Lambda}_{(s)} = m_{D_{(s)}} - m_c + \frac{1}{m_c} \left ( \kappa_1 (1) + 3 \kappa_2 (1) \right ),
\end{eqnarray}
for the $j^P_l = \frac{1}{2}^-$ heavy mesons.
Similarly, from Eqs.(\ref{B0})-(\ref{D0D1p}), we have
\begin{eqnarray}
& & \kappa^{1/2}_1(1) = \frac{m_b m_c}{m_b - m_c} ( \bar{m}_{B^{1/2}_{(s)}} - \bar{m}_{D^{1/2}_{(s)}} -m_b +m_c),\\
& & \kappa^{1/2}_2(1) = \frac{1}{4} m_c ( m_{D^\prime_{(s)1}} - m_{D^*_{(s)0}}),
\end{eqnarray}
and
\begin{eqnarray}
\bar{\Lambda}^{1/2}_{(s)} = m_{D^*_{(s)0}} - m_c + \frac{1}{m_c} \left ( \kappa^{1/2}_1 (1) + 3 \kappa^{1/2}_2 (1) \right ),
\end{eqnarray}
for the $j^P_l = \frac{1}{2}^+$ heavy mesons.
Moreover, we have
\begin{eqnarray}
& & \kappa^{3/2}_1(1) = \frac{m_b m_c}{m_b - m_c} ( \bar{m}_{B^{3/2}_{(s)}} - \bar{m}_{D^{3/2}_{(s)}} -m_b +m_c),\\
& & \kappa^{3/2}_2(1) = \frac{1}{8} m_c (m_{D^*_{(s)2}}-m_{D_{(s)1}}),
\end{eqnarray}
and
\begin{eqnarray}
\bar{\Lambda}^{3/2}_{(s)} = m_{D_{(s)1}} - m_c + \frac{1}{m_c} \left ( \kappa^{3/2}_1 (1) + 5 \kappa^{3/2}_2 (1) \right ),
\end{eqnarray}
for the $j^P_l = \frac{3}{2}^+$ heavy mesons based on Eqs.(\ref{B1})-(\ref{D1D2}).

\section{Conversion formulae for the form factors with different definitions} \label{ConFor}

In this appendix, we give the relevant formulae used to convert the form factors defined in Ref.\cite{LFQM1,LFQM2,ISGW21,LCSRB1} to meet the current definitions, which can be obtained directly by comparing corresponding hadronic matrix elements.

For the $B_{(s)} \rightarrow D^*_{(s)0}$ decays,
\begin{eqnarray}
& & g_+= \frac{1+r}{2 \sqrt{r} (1+r^2-2 r \omega)} \left [ (1-r)^2 f_0 +2 r (1-\omega)f_+ \right], \\
& & g_-= \frac{1-r}{2 \sqrt{r} (1+r^2-2 r \omega)} \left [ (1+r)^2 f_0 -2 r (1+\omega)f_+ \right],
\end{eqnarray}
where
\begin{eqnarray}
& & r=\frac{m_{D^{**}_{(s)}}}{m_{B_{(s)}}}, \\
& & \omega= \frac{m_{B_{(s)}}^2 +m_{D^{**}_{(s)}}^2 -q^2}{2 m_{B_{(s)}} m_{D^{**}_{(s)}}} = \frac{1+r^2-\frac{q^2}{m_{B_{(s)}}^2}}{2r}.
\end{eqnarray}
For the $B_{(s)} \rightarrow D^\prime_{(s)1}$ decays,
\begin{eqnarray}
& & g_A = \frac{2 \sqrt{r}}{1+r} A, \\
& & g_{V_1} = \frac{1+r}{2 \sqrt{r}} V_1, \\
& & g_{V_2} = - \frac{1}{ \sqrt{r} (1+r) (1+r^2-2 r \omega)} \left[ -2 r (1+r) V_0 + (1+r)^2 V_1 +2 r ( r - \omega) V_2 \right ], \\
& & g_{V_3} = - \frac{\sqrt{r}}{  (1+r) (1+r^2-2 r \omega)} \left[ 2 r (1+r) V_0 - (1+r)^2 V_1 +2 ( 1 - r \omega) V_2 \right ].
\end{eqnarray}
For $B_{(s)} \rightarrow D_{(s)1}$ decays,
\begin{eqnarray}
& & f_A= \frac{2 \sqrt{r}}{1+r} A, \\
& & f_{V_1} = \frac{1+r}{\sqrt{r}} V_1, \\
& & f_{V_2} =  - \frac{1}{ \sqrt{r} (1+r) (1+r^2-2 r \omega)} \left[ -2 r (1+r) V_0 + (1+r)^2 V_1 +2 r ( r - \omega) V_2 \right ], \\
& & f_{V_3} = - \frac{\sqrt{r}}{  (1+r) (1+r^2-2 r \omega)} \left[ 2 r (1+r) V_0 - (1+r)^2 V_1 +2 ( 1 - r \omega) V_2 \right ].
\end{eqnarray}
For the $B_{(s)} \rightarrow D^*_{(s)2}$ decays,
\begin{eqnarray}
& & V= - (m_{B_{(s)}} - m_{D^*_{(s)2}}) h, \\
& & A_1 = - \frac{k}{m_{B_{(s)}} - m_{D^*_{(s)2}}}, \\
& & A_2= (m_{B_{(s)}} - m_{D^*_{(s)2}}) b_+, \\
& & A_0= \frac{m_{B_{(s)}} - m_{D^*_{(s)2}}}{2 m_{D^*_{(s)2}}} A_1 - \frac{m_{B_{(s)}} + m_{D^*_{(s)2}}}{2 m_{D^*_{(s)2}}} A_2 -
\frac{(m_{B_{(s)}} - m_{D^*_{(s)2}})^2}{2 m_{D^*_{(s)2}}} b_-,
\end{eqnarray}
and
\begin{eqnarray}
& & k_V= \frac{2 \sqrt{r}}{1+r} V, \\
& & k_{A_1}=  \frac{1+r}{\sqrt{r}} A_1, \\
& & k_{A_2} = - \frac{1}{ \sqrt{r} (1+r) (1+r^2-2 r \omega)} \left[ -2 r (1+r) A_0 + (1+r)^2 A_1 +2 r ( r - \omega) A_2 \right ], \\
& & k_{A_3} = - \frac{\sqrt{r}}{  (1+r) (1+r^2-2 r \omega)} \left[ 2 r (1+r) A_0 - (1+r)^2 A_1 +2 ( 1 - r \omega) A_2 \right ],\\
& & k_{T_1} = \frac{1+r}{   2 \sqrt{r} (1+r^2-2 r \omega)} \left [ 2 r (1-\omega) T_1+ (1-r)^2 T_2 \right ], \\
& & k_{T_2} = \frac{1-r}{   2 \sqrt{r} (1+r^2-2 r \omega)} \left [ -2 r (1+\omega) T_1+ (1+r)^2 T_2 \right ], \\
& & k_{T_3} = - \frac{2 \sqrt{r}}{  (1-r)^2 (1+r^2-2 r \omega)} \left [ - (1-r^2) T_1 + (1-r^2) T_2 + (1-r^2 - 2 r \omega )T_3 \right ].
\end{eqnarray}

\end{document}